\documentclass[a4paper]{report}
\usepackage[utf8]{inputenc}
\usepackage[T1]{fontenc}
\usepackage{RJournal}
\usepackage{amsmath,amssymb,array}
\usepackage{booktabs}
\providecommand{\tightlist}{%
  \setlength{\itemsep}{0pt}\setlength{\parskip}{0pt}}

\usepackage{longtable}

\makeatletter
  \let\@cite@ofmt\@firstofone
  \def\@biblabel#1{}
  \def\@cite#1#2{{#1\if@tempswa , #2\fi}}
\makeatother
\newlength{\cslhangindent}
\newlength{\csllabelwidth}
\usepackage{calc}

\usepackage{mathrsfs}

\usepackage{seqsplit}
\usepackage{enumitem}
\usepackage{xcolor}
\usepackage{pdflscape}
\newcommand{\ltd}[1]{#1}

\begin{document}
\sectionhead{Contributed research article}
\volume{XX}
\volnumber{YY}
\year{20ZZ}
\month{AAAA}

\begin{article}
  \title{ldmppr: Location Dependent Marked Point Processes in R}

\author{by Lane Drew and Andee Kaplan}

\maketitle

\abstract{%
In this article, we present \pkg{ldmppr}, an R package for estimating, evaluating, simulating from, and visualizing location-dependent marked spatial point processes. To date, it has commonly been assumed that the marks associated with a point process are independent of the locations. However, when dealing with many point processes, such as those arising in forestry applications, the independence assumption proves unreasonable. We introduce a practical framework for generating marked point processes with dependence between the marks and locations. We provide a brief discussion of the theory underpinning our modeling approach and outline the use of the package in a typical scenario involving real data. We highlight the functionality of the package for both generating from and assessing the goodness-of-fit of a given model, enabling users to generate realistic point patterns given a reference pattern or parameter values of interest.
}

\section{Introduction}\label{introduction}

Point process models are a rich and complex class of models encompassing
processes that occur over time, space, and potentially include
additional information about the process in the form of marks. Marks,
which are attributes associated with each point (i.e., size or type),
add another layer of complexity, particularly when they depend on
spatially or temporally varying covariates. The behavior of a point
process is typically characterized by the relationship between points.
For example, the simplest process is a homogeneous Poisson process,
which assumes a constant intensity and no interaction between points and
may be described as complete spatial randomness (CSR), where intensity
describes the expected number of points per unit of area. More
structurally involved processes may exhibit regularity (or inhibition),
where points tend to repel each other, or clustering, where collections
of points tend to occur in proximity to each other within the pattern.
Capturing the dynamics of these processes can be challenging even
without the inclusion of marks, and notably more so when this type of
auxiliary information is present. Consequently, researchers often make
the simplifying assumption that the marks associated with a point
process are independent of the primary process itself. However, this
assumption is often hard to justify from a scientific perspective, as in
the case of trees within a forest where the sizes of the trees (i.e.,
canopy volumes) play a role in their distribution over space, and where
the sizes themselves are likely to depend on location specific
information such as elevation, soil moisture, or sunlight availability,
which vary over space. Marked point processes have been used across a
variety of fields including epidemiology, seismology, criminology,
ecology, and the health sciences. Some examples of marked point
processes include ambulance call locations with call severity and
patient gender as marks \citep{bayisaRegularisedSemiparametricComposite2023},
and locations of crime incidences with type of crime as a mark
\citep{mohlerSelfExcitingPointProcess2011}.

In this paper, we present \CRANpkg{ldmppr} \citep{drewLdmpprEstimateSimulate2025a}, an R package that provides a
suite of tools for working with location dependent marked point
processes characterized by regularity in the point pattern. While a
wealth of R packages exist for working with point processes, such as
\CRANpkg{spatstat} \citep{baddeleySpatstatPackageAnalyzing2005} and
\pkg{ptProcess} \citep{hartePtProcessPackageModelling2010}, their focus
is often broad in scope and they fail to incorporate dependence in the
mark distribution with a high degree of flexibility. In contrast,
\CRANpkg{ldmppr} is designed specifically for working with spatial marked
point processes with dependence between the marks and locations that
demonstrate inhibitory behavior in the spatial pattern. The package is
structured to deliver a straightforward and modular workflow that
simplifies the process of model estimation, evaluation, simulation, and
visualization given a reference dataset and a set of corresponding
covariate surfaces in the form of raster images. In addition to the
included models, users can easily adapt the workflow to their specific
needs by substituting appropriate components, while still taking
advantage of the overall package structure.

While inhibitory marked point processes are commonly modeled using Gibbs
processes, they are known to be computationally expensive and difficult
to evaluate efficiently. Instead, we take a likelihood optimization
approach that is computationally scalable, generally tractable, and
enables a straightforward mechanism for simulating from and evaluating
the goodness-of-fit of a model. To achieve this, we employ the
mechanistic approach for equating a marked spatial point process with a
spatio-temporal point process outlined by
\citet{mollerMechanisticSpatiotemporalPoint2016a}, who demonstrated the utility
of the self-correcting process introduced by
\citet{ishamSelfcorrectingPointProcess1979} for modeling marked point processes
with regularity. We extend this framework to include location dependence
in the mark distribution using a class of flexible non-linear models,
improving the applicability of the original approach to a wider variety
of biologically plausible and scientifically interesting processes.

The remainder of the paper is organized as follows. Section \hyperref[sec:background]{2} provides an overview of the relevant point process
theory and modeling framework underlying \CRANpkg{ldmppr}. Section \hyperref[sec:package]{3} describes the package structure and functionality.
Section \hyperref[sec:application]{4} showcases the standard workflow for the
package on an example dataset, with an emphasis on forestry
applications. Section \hyperref[sec:discussion]{5} concludes with a discussion
of the methodological contributions, strengths, limitations, and
potential directions for future development of the package.

\section{Mathematical background}\label{sec:background}

We begin this section with a brief introduction of the theory of marked
point processes to provide context for our modeling approach. We follow
this with a discussion of our proposed approach for mapping a marked
point process to a marked spatio-temporal process that preserves the
dependence between marks and the process and the self-correcting model
that we employ for point processes characterized by regularity.

\subsection{Marked point processes}\label{mpp}

Marked point processes are an extension of point processes that include
additional point specific information in the form of marks. Marks may
represent continuous or discrete quantities (i.e., sizes or species of
trees in a forest), and may be dependent or independent of the primary
generating process. In a process with dependence, the distribution of
the marks depends on the locations of the points, while independent
marks occur independently of the points themselves. In practice, marked
point processes may be modeled by a joint distribution for the points
and their associated marks, or alternatively by the conditional
distribution of the marks given the locations, often described by their
first and second-order characteristics. The first order characteristic
is the mark intensity function which describes the expected mark value
per unit of area, while the second order characteristic, the mark
correlation function, captures the dependence between marks at different
times or locations \citep{schlatherDetectingDependenceMarks2004}.

For purposes of illustration, we consider a marked spatial point process
over a finite domain \(\mathscr{S} \subset \mathbb{R}^2\) defined as
\(\boldsymbol{\Phi} = \{(\boldsymbol{U}_i, \boldsymbol{M}_i): i = 0, \dots, N\}\)
with a two-dimensional nonnegative mark (though this may be extended to
a higher dimensional mark space with additional real valued or discrete
marks). We assume that the process is characterized by regularity such
that points in the process tend to repel each other resulting in higher
interpoint distances than would be observed in a pattern with CSR. We
define \(\boldsymbol{U}_i \in \mathscr{S}\) to be the spatial location of
the \(i\)-th point and
\(\boldsymbol{M}_i = \left (M_{i1}, M_{i2} \right ) \in \mathbb{R}^{+} \times \mathbb{R}^{+}\)
to be the mark vector associated with the \(i\)-th point, where
\(M_{i1} = M_1(\boldsymbol{U}_i)\) is taken to be a measure of size or age
and \(M_{i2} = M_2(\boldsymbol{U}_i)\) is a secondary mark that may be of
additional scientific interest (i.e., height or dbh in forestry
applications). We allow for the possibility that \(M_{i1} = M_{i2}\),
which may be of interest when utilizing our modeling approach to capture
location dependence in the primary mark distribution. Additionally,
\(\boldsymbol{Z}(\boldsymbol{U}_i)\) represents a vector of topographic
(or location specific) covariates at location \(\boldsymbol{U}_i\). We
assume that the mark vector
\(\boldsymbol{M}= \{\boldsymbol{M}_i: i = 0, \dots, N\}\) is ordered
according to the first mark such that the \(M_{i1}\) are increasingly
ordered continuous random variables with
\(0 \leq M_{01} \leq \cdots \leq M_{N1} \leq \tau < \infty\) and where
\(\tau\) is an unknown upper bound for the marks.

A common approach for modeling a process like \(\boldsymbol{\Phi}\) with
regularity in the spatial pattern, would be to adopt a Gibbs process
model as in \citet{mollerStatisticalInferenceSimulation2003}. However, these
models typically include an intractable normalizing constant that makes
them difficult to estimate efficiently and often rely on computationally
expensive MCMC algorithms. Alternatively, assuming that we have a point
process with a structure like \(\boldsymbol{\Phi}\), we can use a
likelihood based approach that allows for straight forward estimation,
model evaluation, and simulation by extending the mechanistic framework
introduced by \citet{mollerMechanisticSpatiotemporalPoint2016a}. The method
maps a spatial marked point process onto a marked spatio-temporal
process conditional on the history of the process by incorporating
location dependence in the mark distribution and allowing for higher
dimensional marks, as described in the following section.

\subsection{Spatio-temporal process mapping with location dependent marks}\label{stmapping}

While marked point processes naturally capture the relationship between
event locations and their associated attributes, directly modeling these
dependencies can be challenging. By mapping a marked spatial point
process onto a spatio-temporal process, we introduce a structured
mechanism for estimation that preserves the dependence between marks and
locations while leveraging established likelihood-based methods for
estimation. Spatio-temporal point processes are a class of models that
represent the occurrences of random events over time and space and may
be extended to include additional marks that depend on space, time, or
both domains simultaneously \citep{RathbunStephenL.1994ASSP}. These processes are driven by a conditional intensity function
\[\lambda^*(t, \boldsymbol{y} \mid \mathscr{F}_t), \quad t > 0, \ \boldsymbol{y} \in \mathscr{S},\]
where \(\mathscr{F}_t\) is the \(\sigma\)-algebra generated by the oldest
point in the process \(\boldsymbol{M}_N\), which we define as the anchor
point for the process, and the points in the process
\((T_i, \boldsymbol{Y}_i)\) with \(T_i < t\) (a more thorough discussion may
be found in \citet{DaleyDarylJ1988AItt}). The conditional intensity function
describes the instantaneous expected rate of events at time \(t\) and at
location \(\boldsymbol{y}\), conditional upon all of the points in the
process occurring before time \(t\).

\ltd{This process may be described as a mechanistic model
\citep{diggleStatisticalAnalysisSpatial2013}, and for a marked point process
of the form of \(\boldsymbol{\Phi}\), defined in Section \hyperref[mpp]{2.1}, we
can equate the process with an infinite marked spatio-temporal point
process
\(\{(T_1,\boldsymbol{Y}_1, X_1),  (T_2,\boldsymbol{Y}_2, X_2), \dots \}\),
where \(\mathscr{T} \subset \mathbb{R}\) is the temporal domain. To facilitate this transformation, we introduce a mapping from the original
marked point process notation \(\boldsymbol{\Phi} = \{(\boldsymbol{U}_i, \boldsymbol{M}_i): i = 0, \dots, N\}\)
to a spatio-temporal representation \(\{(T_i, \boldsymbol{Y}_i, X_i)\}\).
First, we associate each point \((\boldsymbol{U}_i,\boldsymbol{M}_i)\) with a derived time
\(\widetilde{T}_i \stackrel{\tiny def}{=}T(M_{i1}) \in \mathscr{T}\) via
\begin{equation}
\widetilde{T}_i
= 1 - \left [\frac{M_{i1} - \min_{j} (M_{j1})}{\max_{j}(M_{j1}) - \min_{j}(M_{j1})} \right ]^\delta,
\qquad i = 0,\dots,N,
  \label{eq:sizetimemapping}
\end{equation}
where \(\delta\) controls the shape of the mapping function and \(\delta=1\) is a 1-to-1 linear mapping.
This construction yields \(\min_i \widetilde{T}_i = 0\) and \(\max_i \widetilde{T}_i = 1\), so the observed times occur on the interval \([0,1]\), which standardizes the times and ensures comparability across datasets for the temporal parameters while reducing sensitivity to the scale of the mark variable chosen for the temporal mapping.}

\ltd{We then relabel points according to their derived times, so that the spatio-temporal index corresponds to the ordered event times.
Let \(\pi\) be a permutation of \(\{0,\dots,N\}\) such that
\(\widetilde{T}_{\pi(1)} < \widetilde{T}_{\pi(2)} < \cdots < \widetilde{T}_{\pi(N+1)}\).
Define, for \(i = 1,\dots,N+1\),
\[
T_i \stackrel{\tiny def}{=}\widetilde{T}_{\pi(i)}, \qquad \boldsymbol{Y}_i \stackrel{\tiny def}{=}\boldsymbol{U}_{\pi(i)}, \qquad X_i \stackrel{\tiny def}{=}M_{\pi(i),2}.
\]
Thus, \(\boldsymbol{Y}_i \in \mathscr{S}\) is the spatial location of the \(i\)-th event in time, and \(X_i\) is the associated secondary mark, where \(X_i\) may depend on \(\boldsymbol{Y}_i\) and \(\boldsymbol{Z}(\boldsymbol{Y}_i)\).
Under Equation \eqref{eq:sizetimemapping} (for \(\delta > 0\)), larger primary marks correspond to smaller derived times, so ordering by \(T_i\) corresponds to processing points from largest to smallest \(M_{i1}\), consistent with conditioning on the anchor point \((\boldsymbol{U}_N, \boldsymbol{M}_N)\). This representation allows us to generate a point pattern conditional on \((\boldsymbol{U}_N, \boldsymbol{M}_N)\), i.e., the point with the largest primary mark, where the behavior of the process after the final observed event time is ignored as we exploit the relationship between the observed finite processes.}

We define the conditional intensity function of the our marked
spatio-temporal process as follows \[
\lambda^*(t, \boldsymbol{y}, x \mid \mathscr{F}_t), \quad t > 0, \ \boldsymbol{y} \in \mathscr{S}, \ x \geq 0,
\] such that the intensity depends on the points arriving in the process
before time \(t\). The notation \(\lambda^*\) signifies the dependence on
the history of the process, \(\mathscr{F}_t\), which is suppressed in the
notation going forward for concision. We note that under the assumption
of conditional independence between the arrival times and spatial
locations, the conditional intensity function may be decomposed as
follows \[
\lambda^*(t, \boldsymbol{y}, x) = \lambda^*(t) h_t^*(\boldsymbol{y}, x),\quad t > 0, \ \boldsymbol{y} \in \mathscr{S}, \ x \geq 0,
\] where \(\lambda^*(t)\) is the temporal intensity and
\(h_t^*(\boldsymbol{y}, x)\) is the likelihood for the marked spatial
process. We note that \(h_t^*(\boldsymbol{y}, x)\) may be decomposed
further if we assume conditional independence between the locations and
the secondary mark characteristic. This will extend the framework of
\citet{mollerMechanisticSpatiotemporalPoint2016a} to allow for spatio-temporal
dependence in the secondary mark distribution and enable flexible
estimation of the conditional mark distribution incorporating
potentially complex location dependence.

We specify a general parametric model for the conditional intensity,
i.e., \(\lambda^*_{\boldsymbol{\theta}}(t, \boldsymbol{y}, x)\), where
\(\boldsymbol{\theta}\) is an unknown set of parameters and obtain the
spatio-temporal log-likelihood conditional on
\((\boldsymbol{U}_N, \boldsymbol{M}_N) = (\boldsymbol{u}_n, \boldsymbol{m}_n)\)
such that
\[
L(\boldsymbol{\theta}) = \sum_{i=1}^n \log \lambda^*_{\boldsymbol{\theta}}(t_i, \boldsymbol{y}_i, x_i) - \iiint_{(0, t_n) \times \mathscr{S} \times \mathbb{R}^+}  \lambda^*_{\boldsymbol{\theta}}(t, \boldsymbol{y}, x) \mathop{dt}\mathop{d\boldsymbol{y}}\mathop {dx}.
\]
To facilitate model fitting and simulation, we introduce the
temporal-integrated intensity function
\begin{equation}
\Lambda^*(t) = \int_{0}^t \lambda^*(s) \mathop{ds}, \quad t > 0,
  \label{eq:temporalintegratedintensity}
\end{equation}
such that \(S_i = \Lambda^*(T_i), \ i = 1,2,\dots\) form a
unit rate Poisson process on \((0, \infty)\). This relationship allows us
to easily simulate from the process once the model is estimated, which
is a key component in assessing the goodness-of-fit of the estimated
process. Details for the simulation algorithm are provided in Section \hyperref[simulation]{2.5}.

\subsection{Model exhibiting regularity}\label{selfcorrecting}

The framework we have introduced in this section applies to a general
marked point process. We now turn our attention to processes exhibiting
regularity in their spatial pattern as described in Section \hyperref[mpp]{2.1}.
Point processes may be described as self-correcting when the intensity
function decreases as the number of events increases, resulting in an
inhibitory effect that counterbalances event clustering and enforces
regularity in the spatial pattern over time. We define a modified
version of the self-correcting model introduced by
\citet{ishamSelfcorrectingPointProcess1979} with the following intensity
function \[
\lambda^*(t, \boldsymbol{y}, x) = \lambda_{\boldsymbol{\theta}_1}^*(t) h_{\boldsymbol{\theta}_2, t}^*(\boldsymbol{y}) g_{\boldsymbol{\theta}_3, t, \boldsymbol{y}, \boldsymbol{z}(\boldsymbol{y})}^*(x) f_{\boldsymbol{\theta}_4}^*(t, \boldsymbol{y}).
\] This model incorporates an arrival time process
\((\lambda_{\boldsymbol{\theta}_1}^*(t))\), spatial process
\((h_{\boldsymbol{\theta}_2, t}^*(\boldsymbol{y}))\), conditional mark
process \((g_{\boldsymbol{\theta}_3, t, \boldsymbol{y}}^*(x))\), and
spatio-temporal interaction
\((f_{\boldsymbol{\theta}_4}^*(t, \boldsymbol{y}))\). We discuss the
formulations for each component as follows.

\subsubsection{Arrival time process}\label{arrival-time-process}

The arrival time process \(T\) is modeled with intensity function
\(\lambda_{\boldsymbol{\theta}_1}^*(t )\), following
\citet{mollerMechanisticSpatiotemporalPoint2016a}, given by \[
\lambda_{\boldsymbol{\theta}_1}^*(t) =  \exp(\alpha_1 +  \beta_1 t - \gamma_1 N(t)),
\] where \(\boldsymbol{\theta}_1 = (\alpha_1, \beta_1, \gamma_1)\) such
that \(\alpha_1 \in \mathbb{R}\) is a baseline rate,
\(\beta_1 \in [0, \infty)\) is a log-linear function of \(t\), and
\(\gamma_1 \in [0, \infty)\) is the scaling factor for \(N(t)\) where
\(N(t) = \sum_{i = 1}^N\mathbb{I}\{i \geq 0: t_i < t \}\).

\subsubsection{Spatial process}\label{spatial-process}

We model the spatial process with interaction function \(\psi(r)\) such
that \[
\psi_{\boldsymbol{\theta}_2}(r) = \mathbb{I}[r \leq \alpha_2] (r/\alpha_2)^{\beta_2} + \mathbb{I}[r > \alpha_2], \quad r\geq 0,
\] where \(r\) is the Euclidean distance between two points (i.e.,
\(r_{ij} = \lVert \boldsymbol{y}_i - \boldsymbol{y}_j \rVert\)). This
formulation is analogous to the pairwise interaction function in
\citet{diggleMonteCarloMethods1984}. The density for the process is given by
\begin{equation}
h_{\boldsymbol{\theta}_2,i}^* (\boldsymbol{y}_i) = \frac{1}{c_{\boldsymbol{\theta}_2,i}^*} \prod_{j: j < i} \psi_{\boldsymbol{\theta}_2}(r_{ij}),
  \label{eq:spatialdensity}
\end{equation} where \(c_{\boldsymbol{\theta}_2,i}^*\) is the normalizing
constant for the spatial density, which is defined as \[
c_{\boldsymbol{\theta}_2,i}^* = \int_{\mathscr{S}} \prod_{j: j < i} \psi_{\boldsymbol{\theta}_2}(r_{ij}) \mathop{d\boldsymbol{y}},
\] and \(\boldsymbol{\theta}_2 = (\alpha_2, \beta_2 ) \in [0, \infty)^2\).
The spatial density \(h_{\boldsymbol{\theta}_2,i}^*\) defines an
inhibitive circular region around each larger point \(\boldsymbol{y}_j\),
for all points with \(t_j < t_i\), such that the strength of the
interaction diminishes at a polynomial rate for interpoint distances
less than \(\alpha_2\) and disappears for distances greater than
\(\alpha_2\).

\subsubsection{Conditional mark process}\label{conditional-mark-process}

We model the conditional mark process using a flexible non-linear model
such that \[
g_{\boldsymbol{\theta}_3}^*(x_i \mid t_i, \boldsymbol{y}_i, \boldsymbol{z}(\boldsymbol{y}_i)) = \text{G}_{x_i}(t_i, \boldsymbol{y}_i, \boldsymbol{z}(\boldsymbol{y}_i)),
\] where \(\text{G}\) is trained on the feature set
\(\{T_i, \boldsymbol{Y}_i, \boldsymbol{Z}(\boldsymbol{Y}_i)\}\), and
\(\boldsymbol{Z}(\boldsymbol{Y}_i)\) is the set of covariate values at the
location \(\boldsymbol{Y}_i\). We specify this component in generality,
and note that any suitably flexible model may be chosen to model the
conditional mark process (e.g., a random forest
\citep{breimanRandomForests2001} or gradient boosted tree model
\citep{chenXGBoostScalableTree2016}). We assume that the mark process is
conditionally independent given the times and locations and is
parameterized by a non-overlapping set of parameters
\(\boldsymbol{\theta}_3\) when estimating the model. For example, in a
forestry application, assuming conditional independence of marks (i.e.,
sizes) given spatial and temporal covariates implies that tree size at a
given location and time is determined by local environmental factors and
possible interpoint competition indices.

\subsubsection{Spatio-temporal interaction}\label{spatio-temporal-interaction}

The final component of the model is the spatio-temporal interaction
term, as introduced by \citet{mollerMechanisticSpatiotemporalPoint2016a}, which
we define as \[
f_{\boldsymbol{\theta}_4}^*(t_i, \boldsymbol{y}_i) = \exp \left ( -\alpha_4 \sum_{j: j < i} \mathbb{I}[ \Vert \boldsymbol{y}_j - \boldsymbol{y}_i \Vert \leq \beta_4, t_i - t_j \geq \gamma_4] \right ),
\] where
\(\boldsymbol{\theta}_4 = (\alpha_4, \beta_4, \gamma_4) \in [0, \infty)^3\)
with \(\beta_4 = \gamma_4 = 0\) if \(\alpha_4 = 0\). This term connects the
temporal and spatial process components of the full model and captures
the dependence between these components. The form of the interaction
term regulates the development of the process by limiting the occurrence
of points that arrive in close proximity to each other in both space and
time.

\subsection{Parameter estimation}\label{parameter-estimation}

We now consider the procedure for estimating the parameters of the
marked spatio-temporal process. We leverage the separability of the full
log-likelihood under the assumption of conditional independence of the
marks given the locations and arrival times and estimate the conditional
mark process and the spatio-temporal process individually. This approach
allows us to take advantage of highly flexible non-linear machine
learning models to capture the conditional mark process, while
maintaining the computational efficiency of maximum likelihood
estimation for the spatio-temporal process. We note that both processes
depend on the mapping from the original mark space to arrival times,
which is controlled by the parameter \(\delta\) in Equation
\eqref{eq:sizetimemapping}. However, we do not necessarily require the
mappings to be identical for both processes, as the conditional mark
process may exhibit different behavior than the spatial process. We
provide an overview of the estimation procedure for the full model in
practice in Section \hyperref[sec:package]{3}.

For the self-correcting model described in Section \hyperref[selfcorrecting]{2.3}, we estimate the parameters of the process by
optimizing the spatio-temporal component of the log-likelihood, which
may be expressed as \begin{align*}
L(\boldsymbol{\theta_1}, \boldsymbol{\theta_2}, \boldsymbol{\theta_4}) &\propto \sum_{i=1}^n \left[ \log \lambda_{\boldsymbol{\theta}_1}^*(t_i) + \log h_{\boldsymbol{\theta}_2,i}^*(\boldsymbol{y}_i) + \log f_{\boldsymbol{\theta}_4}^*(t_i, \boldsymbol{y}_i) \right] \\
&\qquad - \iint_{(0, t_n) \times \mathscr{S}} \lambda_{\boldsymbol{\theta}_1}^*(t) h_{\boldsymbol{\theta}_2,t}^*(\boldsymbol{y}) f_{\boldsymbol{\theta}_4}^*(t, \boldsymbol{y}) \, dt \, d\boldsymbol{y}.
\end{align*} The parameter sets
\(\boldsymbol{\theta}_1, \boldsymbol{\theta}_2, \boldsymbol{\theta}_4\)
can be estimated using any suitable optimization algorithm that allows
for bound constraints on the parameter space.

For the conditional mark process, we estimate the parameter set
\(\boldsymbol{\theta}_3\) using a flexible model that allows for the
incorporation of spatial and temporal covariates. The set of features
used in training the model which correspond to location specific
covariates \(\boldsymbol{Z}(\boldsymbol{Y}_i)\) may be derived from raster
images or other spatial data sources and may include interpoint
competition metrics, as discussed in Section \hyperref[markmodel]{3.3}.

We combine the estimated parameter values for the conditional mark
process and the spatio-temporal process to obtain the full set of
estimated parameters for the marked spatio-temporal process. This
approach is modular and allows for a wide variety of models to be
substituted for the conditional mark process. In theory, this framework
also holds for alternative spatio-temporal processes, such as those
exhibiting clustering or other spatial patterns, though the specific
form of the likelihood components would need to be adjusted accordingly.

\subsection{Simulating from the spatio-temporal process}\label{simulation}

Utilizing the framework outlined in this section and estimates of the
parameter sets
\(\boldsymbol{\theta}_1\), \(\boldsymbol{\theta}_2\), \(\boldsymbol{\theta}_3\), \(\boldsymbol{\theta}_4\),
we can easily generate realizations from the marked spatio-temporal
point process on \([0, 1] \times \mathscr{S}\). Given
\(S_i = \Lambda^*(T_i), \ i = 1,2,\dots\) as defined in Equation
\eqref{eq:temporalintegratedintensity}, we have that \begin{equation*}
  T_1 = \frac{1}{\beta_1} \log \{ 1 + \beta_1(\gamma_1 - \alpha_1) S_1 \}
\end{equation*} and for \(i = 2, 3, \dots\) \begin{align*}
  T_i &= \frac{1}{\beta_1} \log \biggl [ \exp(\beta_1 T_{i-1}) + \beta_1 \exp(\gamma_1 i - \alpha_1) S_i \\
  &\qquad - \sum_{j=1}^{i-1} \{\gamma_1(i - j)\} \{ \exp(\beta_1 T_j) - \exp(\beta_1 T_{j-1}) \}\biggr ]
\end{align*} such that we can generate a realization \(t_1 < \dots < t_n\)
under the self-correcting model on \([0, 1]\) as follows.

\begin{enumerate}
\def\labelenumi{\arabic{enumi}.}
\tightlist
\item
  Generate a draw from a Poisson distribution with rate
  \(\rho_{\boldsymbol{\widehat{\theta}}_1}^*(t) = \exp \{\widehat{\alpha}_1 + \widehat{\beta}_1 t_n \}\),
  where \(t_n = 1\) to obtain the number of possible arrival times
  \(n^*\).
\item
  For \(i = 1, \dots, n^*\), generate a draw from the uniform
  distribution on \((0, t_n)\).
\item
  Order the uniform draws such that \(u_1 < \dots < u_{n^*}\) to obtain
  the candidate arrival times \(t_1^*,\dots, t_{n^*}^*\).
\item
  For \(i = 1, \dots, n^*\), calculate the acceptance ratio \[
    r_i^* = \frac{\lambda_{\boldsymbol{\widehat{\theta}}_1}^*(t_i^*)}{\exp \{\widehat{\alpha}_1 + \widehat{\beta}_1 t_n \}},
    \] then generate a draw \(u_i^*\) from the uniform distribution on
  \((0, 1)\) and accept \(t_i^*\) if \(u_i^* < r_i^*\).
\item
  Given the set of accepted arrival times \(\{t_i\}_{i=1}^n\), generate
  points \(\boldsymbol{y}_i\) from the density
  \(h_{\boldsymbol{\widehat{\theta}}_2, i}^*(\boldsymbol{y}_i)\) using
  rejection sampling with acceptance probability derived from the
  unnormalized component of the spatial interaction in Equation
  \eqref{eq:spatialdensity} and a uniform proposal distribution over
  \(\mathscr{S}\).
\item
  Given the set of \(\{t_i, \boldsymbol{y}_i\}_{i=1}^n\), thin the
  process such that each point is kept with probability
  \(f_{\boldsymbol{\widehat{\theta}}_4}^*(t_i, \boldsymbol{y}_i)\) and the
  kept points are a realization of the spatio-temporal process.
\item
  For \(i = 1, \dots, n\), given
  \((t_i, \boldsymbol{y}_i, \boldsymbol{z}(\boldsymbol{y}_i))\),
  generate \(x_i\) from the density
  \(g_{\boldsymbol{\widehat{\theta}}_3}^*(x_i \mid t_i, \boldsymbol{y}_i, \boldsymbol{z}(\boldsymbol{y}_i))\).
\end{enumerate}

This modeling approach allows us to employ interpretable and
computationally tractable likelihood methods for estimating and
simulating from the spatio-temporal process and provides a high degree
of flexibility when selecting and training a model for the conditional
mark distribution as discussed in the following section.

\section{Package structure}\label{sec:package}

In this section, we introduce the core functionality for the
\CRANpkg{ldmppr} package and detail the key functions associated with
spatio-temporal process and mark model estimation, model goodness-of-fit
evaluation, simulation, and visualization. We note that the package is
currently designed for working with marked point processes that can be
mapped onto spatio-temporal processes where the pattern is characterized
by regularity and the mark distribution is dependent on location
specific covariate information.

\subsection{Workflow}\label{workflow}

We begin by describing the standard workflow that we envision for using
the package. We decompose the task of working with marked point
processes into a handful of straightforward and manageable steps
supported by an intuitive set of functions. \ltd{The process is outlined
as follows, where the output of each step can be passed as the input to successive stages where appropriate.}

\begin{enumerate}
\def\labelenumi{\arabic{enumi}.}
\tightlist
\item
  Estimate the parameters of a self-correcting point process given a
  reference dataset.
\item
  Train a mark model given the reference data set and topographic
  covariate surfaces in the form of rasters.
\item
  Check the fit of the estimated model using various non-parametric
  summaries for point processes and global envelope tests.
\item
  Simulate and visualize datasets from the fitted model.
\end{enumerate}

We anticipate that users of the package will have a point process that
they are interested in investigating, and we provide the tools to
facilitate that exploration. We also note the modular structure of the
package such that a user may provide their own estimated mark model in
lieu of one of the currently available options in the package. In the
remaining portions of this section, we detail the key functionality for
each step given above.

\subsection{Self-correcting model estimation}\label{scestimation}

\ltd{The first step in the model estimation procedure is to estimate the
parameters of the self-correcting model detailed in Section \hyperref[selfcorrecting]{2.3} that captures the spatio-temporal process in the
data. Given a reference dataset, we must define a mapping between our
initial mark (i.e., size) and the arrival time in the process by specifying a value for \(\delta\). This allows the user to
specify how the marks are mapped to arrival times using the
transformation given in Equation \eqref{eq:sizetimemapping}. Once a mapping is
chosen, the user should select an optimization strategy (\texttt{“local”}, \texttt{“global\_local”}, or \texttt{“multires\_global\_local”}) that determines how the optimization is performed. For the \texttt{"local"} strategy, the optimization is run using a single layer with a local algorithm. For the \texttt{“global\_local”} strategy, the optimization is run using two layers that starts with a global optimization that is then updated in a local sweep. For the \texttt{“multires\_global\_local”} strategy, this extends the \texttt{“global\_local”} approach to allow users to further refine the second stage local estimate at higher resolution. Once a strategy has been selected, users must specify both the integration grid and the optimization budget using \texttt{ldmppr\_grids()} and \texttt{ldmppr\_budgets()}, respectively. The grid controls the numerical resolution used to approximate the likelihood integral at each fitting level, while the budget controls how much optimization effort is allocated to each stage (global search, first-level local search, and refinement-level local search), including stopping criteria such as evaluation limits and convergence tolerances. In this context, ``budget'' refers to computational effort, not model complexity. With both components defined, parameter estimation for the self-correcting process proceeds via \texttt{estimate\_process\_parameters()}, where the arguments of the function are defined as follows.}

\renewcommand{\arraystretch}{1.1}
\begin{longtable}{>{\raggedleft\arraybackslash}p{0.26\linewidth}p{0.68\linewidth}}
\toprule
Argument & Description\\
\midrule
\texttt{\detokenize{data}} & a data.frame or matrix. Must contain either columns \texttt{\detokenize{(time, x, y)}} or \texttt{\detokenize{(x, y, size)}}. If a matrix is provided without time, it must have  column names \texttt{\detokenize{c(``x'',``y'',``size'')}}.\\
\texttt{\detokenize{process}} & type of process used (currently supports \texttt{\detokenize{``self_correcting''}}).\\
\texttt{\detokenize{grids}} & a \texttt{\detokenize{ldmppr_grids}} object specifying the integration grid schedule (single-level or multi-resolution). The integration bounds are taken from \texttt{\detokenize{grids$upper_bounds}}.\\
\texttt{\detokenize{budgets}} & a \texttt{\detokenize{ldmppr_budgets}} object controlling optimizer options for the global stage and local stages (first level vs refinement levels).\\
\texttt{\detokenize{parameter_inits}} & (optional) numeric vector of length 8 giving initialization values for the model parameters. If \texttt{\detokenize{NULL}}, defaults are derived from \texttt{\detokenize{data}} and \texttt{\detokenize{grids$upper_bounds}}.\\
\addlinespace
\texttt{\detokenize{delta}} & (optional) numeric scalar or vector. Used only when \texttt{\detokenize{data}} does not contain \texttt{\detokenize{time}} (i.e., data has \texttt{\detokenize{(x,y,size)}}).
\begin{itemize}
  \item If \texttt{\detokenize{length(delta) == 1}}, fits the model once using \texttt{\detokenize{power_law_mapping(size, delta)}}.
  \item If \texttt{\detokenize{length(delta) > 1}}, performs a delta-search by fitting the model for each candidate value and selecting the best objective. If \texttt{\detokenize{refine_best_delta = TRUE}} and multiple grid levels are used, the best delta is refined on the remaining (finer) grid levels.
\end{itemize}
If \texttt{\detokenize{data}} already contains \texttt{\detokenize{time}}, \texttt{\detokenize{delta}} is ignored when \texttt{\detokenize{length(delta)==1}} and is an error when \texttt{\detokenize{length(delta)>1}}.\\
\texttt{\detokenize{parallel}} & \texttt{\detokenize{TRUE}} or \texttt{\detokenize{FALSE}} specifying \CRANpkg{furrr}/\CRANpkg{future} to parallelize either:
\begin{enumerate}[label=\alph*)]
  \item over candidate \texttt{\detokenize{delta}} values (when \texttt{\detokenize{length(delta) > 1}}), and/or
  \item over local multi-start initializations (when \texttt{\detokenize{starts$local > 1}}), and/or
  \item over global restarts (when \texttt{\detokenize{starts$global > 1}}).
\end{enumerate}\\
\texttt{\detokenize{num_cores}} & number of workers to use when \texttt{\detokenize{set_future_plan = TRUE}}.\\
\texttt{\detokenize{set_future_plan}} & \texttt{\detokenize{TRUE}} or \texttt{\detokenize{FALSE}}, temporarily sets \texttt{\detokenize{future::plan(multisession, workers = num_cores)}} and restores the original plan on exit.\\
\texttt{\detokenize{strategy}} & character string specifying the estimation strategy:
\begin{itemize}
  \item \texttt{\detokenize{``local''}}: local optimization only (single-level or multi-level polish).
  \item \texttt{\detokenize{``global_local''}}: global optimization then local polish (single grid level).
  \item \texttt{\detokenize{``multires_global_local''}}: multi-resolution (coarsest uses global+local; refinements use local only).
\end{itemize}\\
\addlinespace
\texttt{\detokenize{global_algorithm, 
 local_algorithm}} & NLopt algorithms to use for the global and local optimization stages, respectively.\\
\texttt{\detokenize{starts}} & a list controlling restart and jitter behavior:
\begin{itemize}
  \item \texttt{\detokenize{global}}: integer, number of global restarts at the first/coarsest level (default 1).
  \item \texttt{\detokenize{local}}: integer, number of local multi-starts per level (default 1).
  \item \texttt{\detokenize{jitter_sd}}: numeric SD for jittering (default 0.35).
  \item \texttt{\detokenize{seed}}: integer base seed (default 1).
\end{itemize}\\
\texttt{\detokenize{finite_bounds}} & (optional) list with components \texttt{\detokenize{lb}} and \texttt{\detokenize{ub}} giving finite lower and upper bounds for all 8 parameters. If \texttt{\detokenize{NULL}}, bounds are derived from \texttt{\detokenize{parameter_inits}}. Global algorithms and select local algorithms in NLopt require finite bounds.\\
\texttt{\detokenize{refine_best_delta}} & \texttt{\detokenize{TRUE}} or \texttt{\detokenize{FALSE}}. If \texttt{\detokenize{TRUE}} and \texttt{\detokenize{length(delta) > 1}}, performs refinement of the best \texttt{\detokenize{delta}} across additional grid levels (if available).\\
\texttt{\detokenize{rescore_control}} & controls candidate rescoring and bound-handling behavior in multi-resolution fitting. Can be either:
\begin{itemize}
  \item a single logical value (toggle rescoring on/off while keeping defaults), or
  \item a named list with any of: \texttt{\detokenize{enabled}}, \texttt{\detokenize{top}}, \texttt{\detokenize{objective_tol}}, \texttt{\detokenize{param_tol}}, \texttt{\detokenize{avoid_bound_solutions}}, \texttt{\detokenize{bound_eps}}.
\end{itemize}
Defaults are \texttt{\detokenize{list(enabled = TRUE, top = 5L, objective_tol = 1e-6,
                    param_tol = 0.10, avoid_bound_solutions = TRUE,
                    bound_eps = 1e-8)}}.\\
\addlinespace
\texttt{\detokenize{verbose}} & \texttt{\detokenize{TRUE}} or \texttt{\detokenize{FALSE}} indicating whether to show progress of model estimation.\\
\bottomrule
\end{longtable}

\ltd{This function makes use of the R implementation of the NLopt library \citep{johnsonNLoptNonlinearoptimizationPackage2008} and returns an \texttt{ldmppr\_fit} object containing the estimated parameters and optimization details. The optimization procedure is designed to be flexible and allows for a variety of strategies and algorithm choices and combinations through the \texttt{nloptr} engine. The optimal choice of strategy and global/local algorithm may depend on the dataset, but the recommended default is a \texttt{“global\_local”} strategy using the \texttt{NLOPT\_GN\_CRS2\_LM} global algorithm paired with the \texttt{NLOPT\_LN\_BOBYQA} local algorithm. The \texttt{NLOPT\_GN\_CRS2\_LM} algorithm is a global optimization algorithm that implements the ``controlled random search'' method, originally introduced by \citet{priceGlobalOptimizationControlled1983}, with the ``local mutation'' modification defined by \citet{kaeloVariantsControlledRandom2006}. This algorithm is designed to be robust and efficient for global optimization problems, as it is less dependent on the initialization values and has been found to perform well in our testing. The recommended local optimization algorithm \texttt{NLOPT\_LN\_BOBYQA} is a derivative-free optimization algorithm that is designed for bound-constrained optimization problems through iteratively building a quadratic approximation of the objective function \citep{powell2009}. Another local optimization option is \texttt{NLOPT\_LN\_SBPLX}, which is an implementation of the ``Subplex'' algorithm introduced by \citet{rowanFunctionalStabilityAnalysis1990} that incorporates explicit bound constraints. This algorithm is a more efficient and robust implementation of the original Nelder-Mead algorithm \citep{nelderSimplexMethodFunction1965}. If the user wishes to test a set of different \(\delta\) values for the mapping function, the \texttt{delta} argument may be specified as a vector and the \texttt{refine\_best\_delta} argument enables refinement of the optimal \(\delta\) from the initial stage in additional layers.}

\ltd{In practice, we recommend a staged setup for \texttt{estimate\_process\_parameters()}. First, set \texttt{delta\ =\ 1} and obtain a baseline fit with a coarse grid and moderate optimization budgets. If diagnostics suggest poor fit, then increase optimization effort by adding additional starts and moving to the \texttt{“multires\_global\_local”} strategy with stricter tolerances. If users wish to tune \texttt{delta}, in the absence of prior knowledge of the relationship, we recommend searching a small neighborhood around 1 (rather than a wide range), selecting the best objective, and then refining only that candidate with finer grids. This strategy typically improves robustness while keeping computational cost manageable and allows users to refine the fit as needed based on the available diagnostics.}

\ltd{In addition to the optimization strategy, the user may also specify the number of global and local multi-starts and the amount of jittering to add to the initial parameter values for each start within the \texttt{starts} argument. Jittering multiple starts can help to improve the optimization results, and reduce dependence on the initial values. This will increase the computational time for estimation, though this can be offset by parallelization over the multi-starts using the \texttt{parallel} and \texttt{num\_cores} arguments. The \texttt{estimate\_process\_parameters()} function is constructed to derive reasonable initial values and bound constraints based on the provided reference dataset, and we encourage users to use these defaults unless they have substantial knowledge about the reasonable values and parameter bounds for their specific application. We also note
that the success of the optimization procedure may depend on the number of function evaluations and tolerance settings for the various algorithms, which can be controlled through the \texttt{budgets} argument. We recommend starting with a moderate number of function evaluations (e.g., 1000 for the global stage and 500 for the local stage) and adjusting as needed based on the results and computational resources available. The choice of integration grid granularity, specified through \texttt{grids}, for the spatial and temporal components should reflect the scale of the data and the expected range of interactions. See Section \hyperref[sec:application]{4} for an example of the effect of the choice of strategy, grids, and budgets on the model fit.}

\subsection{Mark model training}\label{markmodel}

The second step in the model estimation procedure is to train the
conditional mark model. We use the reference data with the mapped
arrival times derived in the self-correcting model estimation step, or
an alternate mapping, in concert with a set of covariate surfaces in the
form of raster images. The raster images may be pre-processed using the
\texttt{scale\_rasters()} function, or may be provided in their raw form. To
train the model, we use the \texttt{train\_mark\_model()} function, where the
arguments of the function are provided as follows.

\renewcommand{\arraystretch}{1.1}
\begin{longtable}{>{\raggedleft\arraybackslash}p{0.26\linewidth}p{0.68\linewidth}}
\toprule
Argument & Description\\
\midrule
\texttt{\detokenize{data}} & a data.frame or a \texttt{\detokenize{ldmppr_fit}} object.\\
\texttt{\detokenize{raster_list}} & a list of raster objects.\\
\texttt{\detokenize{scaled_rasters}} & \texttt{\detokenize{TRUE}} or \texttt{\detokenize{FALSE}} indicating whether the rasters have been scaled.\\
\texttt{\detokenize{model_type}} & the machine learning model type (\texttt{\detokenize{``xgboost''}} or \texttt{\detokenize{``random_forest''}}).\\
\texttt{\detokenize{xy_bounds}} & a vector of domain bounds (2 for x, 2 for y). If \texttt{\detokenize{data}} is an \texttt{\detokenize{ldmppr_fit}} and \texttt{\detokenize{xy_bounds}} is \texttt{\detokenize{NULL}}, defaults to \texttt{\detokenize{c(0, b_x, 0, b_y)}} derived from fit.\\
\addlinespace
\texttt{\detokenize{delta}} & (optional) numeric scalar used only when \texttt{\detokenize{data}} contains \texttt{\detokenize{(x,y,size)}} but not \texttt{\detokenize{time}}. If \texttt{\detokenize{data}} is an \texttt{\detokenize{ldmppr_fit}} and time is missing, the function will infer the \texttt{\detokenize{delta}} value from the fit.\\
\texttt{\detokenize{save_model}} & \texttt{\detokenize{TRUE}} or \texttt{\detokenize{FALSE}} indicating whether to save the generated model.\\
\texttt{\detokenize{save_path}} & the path for saving the generated model.\\
\texttt{\detokenize{parallel}} & \texttt{\detokenize{TRUE}} or \texttt{\detokenize{FALSE}} indicating whether to use parallelization in model training.\\
\texttt{\detokenize{num_cores}} & number of cores to use in parallel model training (if \texttt{\detokenize{parallel}} is \texttt{\detokenize{TRUE}}).\\
\addlinespace
\texttt{\detokenize{include_comp_inds}} & \texttt{\detokenize{TRUE}} or \texttt{\detokenize{FALSE}} indicating whether to generate and use competition indices as covariates.\\
\texttt{\detokenize{competition_radius}} & distance for competition radius if \texttt{\detokenize{include_comp_inds}} is \texttt{\detokenize{TRUE}}.\\
\texttt{\detokenize{edge_correction}} & type of edge correction to apply (\texttt{\detokenize{``none''}}, \texttt{\detokenize{``toroidal''}}, or \texttt{\detokenize{``truncation''}}).\\
\texttt{\detokenize{selection_metric}} & metric to use for identifying the optimal model (\texttt{\detokenize{``rmse''}}, \texttt{\detokenize{``mae''}}, or \texttt{\detokenize{``rsq''}}).\\
\texttt{\detokenize{cv_folds}} & number of cross-validation folds to use in model training. If \texttt{\detokenize{cv_folds <= 1}}, tuning is skipped and the model is fit once with default hyperparameters.\\
\addlinespace
\texttt{\detokenize{tuning_grid_size}} & size of the tuning grid for hyperparameter tuning.\\
\texttt{\detokenize{seed}} & integer seed for reproducible resampling/tuning/model fitting.\\
\texttt{\detokenize{verbose}} & \texttt{\detokenize{TRUE}} or \texttt{\detokenize{FALSE}} indicating whether to show progress of model training.\\
\bottomrule
\end{longtable}

The \texttt{train\_mark\_model()} function allows users to select between a
random forest or gradient boosted tree model, using the \CRANpkg{ranger}
\citep{wrightRangerFastImplementation2017} and \CRANpkg{xgboost}
\citep{chenXgboostExtremeGradient2026} engines respectively, where these
models are effective at capturing potentially complex non-linear
relationships. \ltd{Users choose a model selection criterion}, either root mean
squared error (RMSE) or mean absolute error (MAE), and may employ cross
validation and hyperparameter tuning using a grid design that optimizes
the maximum entropy of the hyperparameter space. \ltd{As a general rule, we recommend that users
start with a moderate number of cross-validation folds (e.g., 5) and a tuning grid size of 100, and adjust as needed based on the results and computational resources available, where parallelization can be used to speed up the tuning process. The \texttt{train\_mark\_model()} function is designed to be flexible and allows users to specify a variety of options for model training, including the ability to save the trained model for future use.}

The function also allows users to incorporate a collection of interpoint competition
metrics at a specified neighborhood size to capture additional trends
that are not accounted for by the topographic covariates. The metrics
include nearest neighbor distance, number of neighbors, average neighbor
distance, nearest neighbor arrival time, sum of neighbor arrival times,
and the ratio of nearest neighbor distance and arrival time. For an in
depth discussion of competition indices and their construction, see
\citet{pommereningTammReviewTree2018} and \citet{contrerasEvaluatingTreeCompetition2011}. Finally, users may select an
edge correction mechanism when training the model to account for the
possibility that the reference dataset provided is a subset of a larger
dataset and that unobserved points are impacting the observed mark
values (i.e., sizes).

\subsection{Goodness-of-fit checks for the fitted model}\label{goodness-of-fit-checks-for-the-fitted-model}

Once the self-correcting model parameters are estimated and the mark
model is trained, we turn our attention to assessing how well the fitted
models capture the dynamics observed in the reference dataset. To
accomplish this, we use the \texttt{check\_model\_fit()} function, which
provides global envelope tests, using the \CRANpkg{GET} package
\citep{myllymakiGlobalEnvelopeTests2017}, for a collection of standard
non-parametric marked point process summary functions. \ltd{We include the \(L\), \(F\), \(G\), \(J\), \(E\), and \(V\) functions as evaluation metrics, together with a combined global envelope test that provides a single Monte Carlo \(p\)-value across the full set of summaries.}

\ltd{The \(L\)-function is a variance-stabilized transformation of Ripley's \(K\)-function and is commonly interpreted as a second-order measure of spatial interaction: relative to CSR, \(L(r)\) above (below) its theoretical reference suggests clustering (inhibition) at spatial scale \(r\) \citep[\citet{diggleStatisticalAnalysisSpatial2013}]{ripleySecondorderAnalysisStationary1976}. The \(F\) and \(G\) functions summarize complementary nearest-neighbor behavior: \(F(r)\) (the empty-space function) describes distances from arbitrary locations in \(\mathscr{S}\) to the nearest event, whereas \(G(r)\) (the nearest-neighbor distribution function) describes distances from events to their nearest neighboring event \citep{diggleStatisticalAnalysisSpatial2013}. The \(J\)-function combines these summaries via \(J(r) = {1-G(r)}/{1-F(r)}\); values \(J(r) > 1\) are typically associated with inhibition/regularity and values \(J(r) < 1\) with clustering \citep{lieshoutJFunctionMarkedPoint2006}. To assess the mark component, we use the mark summary functions \(E\) and \(V\) (implemented as \texttt{Emark} and \texttt{Vmark} in \pkg{spatstat}), which describe how the conditional mean and conditional variability of marks change with interpoint distance, thereby providing diagnostics for mark--location dependence beyond the unmarked spatial structure \citep[\citet{baddeleySpatialPointPatterns2015}]{schlatherDetectingDependenceMarks2004}. Finally, global envelope testing provides a principled way to control multiplicity across distances (and across multiple functions in the combined test) when comparing the observed summaries to simulations from the fitted model \citep{myllymakiGlobalEnvelopeTests2017}.}

The arguments of the \texttt{check\_model\_fit()} function are defined as follows.

\renewcommand{\arraystretch}{1.1}
\begin{longtable}{>{\raggedleft\arraybackslash}p{0.26\linewidth}p{0.68\linewidth}}
\toprule
Argument & Description\\
\midrule
\texttt{\detokenize{reference_data}} & (optional) a marked \texttt{\detokenize{ppp}} object for the reference dataset. If \texttt{\detokenize{NULL}}, the reference pattern is derived from \texttt{\detokenize{process_fit}} when \texttt{\detokenize{process_fit}} is an \texttt{\detokenize{ldmppr_fit}} and contains \texttt{\detokenize{data_original}} (preferred) or \texttt{\detokenize{data}} with columns \texttt{\detokenize{(x, y, size)}}.\\
\texttt{\detokenize{t_min}} & minimum value for time.\\
\texttt{\detokenize{t_max}} & maximum value for time.\\
\texttt{\detokenize{process}} & type of process used (currently supports \texttt{\detokenize{``self_correcting''}}).\\
\texttt{\detokenize{process_fit}} & either an \texttt{\detokenize{ldmppr_fit}} object (from \texttt{\detokenize{estimate_process_parameters}}) or a numeric vector of length 8 giving the process parameters.\\
\addlinespace
\texttt{\detokenize{anchor_point}} & (optional) vector of \texttt{\detokenize{(x, y)}} coordinates of the point to condition on. If \texttt{\detokenize{NULL}}, inferred from the reference data (largest mark if available) or from the \texttt{\detokenize{ldmppr_fit}}.\\
\texttt{\detokenize{raster_list}} & (optional) a list of raster objects used for predicting marks. Required when \texttt{\detokenize{mark_mode = ``mark_model''}} unless rasters are stored in \texttt{\detokenize{mark_model}}.\\
\texttt{\detokenize{scaled_rasters}} & \texttt{\detokenize{TRUE}} or \texttt{\detokenize{FALSE}} indicating whether rasters are already scaled. Ignored when \texttt{\detokenize{mark_mode = ``time_to_size''}}.\\
\texttt{\detokenize{mark_model}} & (optional) a mark model object used when \texttt{\detokenize{mark_mode = ``mark_model''}}. May be an \texttt{\detokenize{ldmppr_mark_model}}, \texttt{\detokenize{model_fit}}, or \texttt{\detokenize{workflow}}.\\
\texttt{\detokenize{xy_bounds}} & (optional) vector of bounds as \texttt{\detokenize{c(a_x, b_x, a_y, b_y)}}. If \texttt{\detokenize{NULL}}, inferred from \texttt{\detokenize{reference_data}}'s window when \texttt{\detokenize{reference_data}} is provided; otherwise inferred from the \texttt{\detokenize{ldmppr_fit}} with lower bounds assumed to be 0.\\
\addlinespace
\texttt{\detokenize{include_comp_inds}} & \texttt{\detokenize{TRUE}} or \texttt{\detokenize{FALSE}} indicating whether to compute competition indices.\\
\texttt{\detokenize{competition_radius}} & distance for competition radius if \texttt{\detokenize{include_comp_inds = TRUE}}.\\
\texttt{\detokenize{thinning}} & \texttt{\detokenize{TRUE}} or \texttt{\detokenize{FALSE}} indicating whether to use the thinned simulated values.\\
\texttt{\detokenize{edge_correction}} & type of edge correction to apply (\texttt{\detokenize{``none''}} or \texttt{\detokenize{``toroidal''}}).\\
\texttt{\detokenize{n_sim}} & number of simulated datasets to generate.\\
\addlinespace
\texttt{\detokenize{save_sims}} & \texttt{\detokenize{TRUE}} or \texttt{\detokenize{FALSE}} indicating whether to save and return the simulated metrics.\\
\texttt{\detokenize{verbose}} & \texttt{\detokenize{TRUE}} or \texttt{\detokenize{FALSE}} indicating whether to show progress of model checking. When \texttt{\detokenize{TRUE}}, progress is reported via \texttt{\detokenize{progressr}} (if available) and is compatible with parallel execution.\\
\texttt{\detokenize{seed}} & integer seed for reproducibility.\\
\texttt{\detokenize{parallel}} & \texttt{\detokenize{TRUE}} or \texttt{\detokenize{FALSE}}. If \texttt{\detokenize{TRUE}}, simulations are run in parallel via \texttt{\detokenize{furrr}}/\texttt{\detokenize{future}}.\\
\texttt{\detokenize{num_cores}} & number of workers to use when \texttt{\detokenize{parallel = TRUE}}. Defaults to one fewer than the number of detected cores.\\
\addlinespace
\texttt{\detokenize{set_future_plan}} & \texttt{\detokenize{TRUE}} or \texttt{\detokenize{FALSE}}. If \texttt{\detokenize{TRUE}} and \texttt{\detokenize{parallel = TRUE}}, set a temporary \texttt{\detokenize{future}} plan internally and restore the previous plan on exit.\\
\texttt{\detokenize{mark_mode}} & (optional) mark generation mode. \texttt{\detokenize{``mark_model''}} uses \texttt{\detokenize{predict()}} on a mark model, while \texttt{\detokenize{``time_to_size''}} maps simulated times back to sizes via \texttt{\detokenize{delta}}. If \texttt{\detokenize{NULL}}, inferred as \texttt{\detokenize{``mark_model''}} when \texttt{\detokenize{mark_model}} is provided, otherwise \texttt{\detokenize{``time_to_size''}}.\\
\texttt{\detokenize{fg_correction}} & correction used for F/G/J summaries (\texttt{\detokenize{``km''}} or \texttt{\detokenize{``rs''}}).\\
\texttt{\detokenize{max_attempts}} & maximum number of simulation attempts when rejection occurs due to non-finite summaries.\\
\bottomrule
\end{longtable}

\ltd{The \texttt{check\_model\_fit()} function allows users to simulate a collection
of datasets using the estimated parameters from the self-correcting
process and trained mark model, which can be passed directly as \texttt{ldmppr\_fit} and \texttt{ldmppr\_mark\_model} objects through the \texttt{process\_fit} and \texttt{mark\_model} arguments respectively. The non-parametric summary functions are
calculated for each dataset and the global envelopes are obtained across
the whole collection of simulated datasets for each metric, as well as a
combined test across all metrics. This allows the user to gauge whether
the realizations from the estimated location dependent marked point
process accurately reflect the dynamics observed in the reference
dataset across a variety of different metrics. Each individual metric
includes a \(p\)-value range indicating the compatibility of the reference
dataset with the simulated datasets, and the combined test provides a
single \(p\)-value across the entire set of metrics allowing users to
quickly gauge the overall fit of the model. Contrary to many testing procedures, a non-significant result (i.e., a large \(p\)-value) indicates that the reference dataset is not incompatible with the fitted model, whereas a significant result (i.e., a small \(p\)-value) indicates that the reference dataset is incompatible with the fitted model. We recommend starting with a moderate number of simulations (e.g., 200 or 500) to balance accuracy of the \(p\)-value estimates with computational time, and adjusting as needed based on the results and resources available. Parallelization can be employed to speed up the simulation and testing process when a larger number of simulations is desired.
For additional discussion of the summary functions included in \texttt{check\_model\_fit()} see \citet{mollerStatisticalInferenceSimulation2003} and
\citet{schlatherDetectingDependenceMarks2004}.}

\subsection{Simulation and visualization}\label{simulation-and-visualization}

When a user is satisfied with the estimated location dependent marked
point process obtained in the model estimation, training, and goodness
of fit checking steps, they can proceed with simulating realizations
from the process and visualizing the results. The \texttt{simulate\_mpp()}
function provides an efficient way to generate a realization from the
process, where the arguments for the function are as follows.

\renewcommand{\arraystretch}{1.1}
\begin{longtable}{>{\raggedleft\arraybackslash}p{0.26\linewidth}p{0.68\linewidth}}
\toprule
Argument & Description\\
\midrule
\texttt{\detokenize{process}} & type of process used (currently supports \texttt{\detokenize{``self_correcting''}}).\\
\texttt{\detokenize{process_fit}} & either (1) an \texttt{\detokenize{ldmppr_fit}} object returned by \texttt{\detokenize{estimate_process_parameters}}, or (2) a numeric vector of length 8 giving self-correcting parameters $(\alpha_1, \beta_1, \gamma_1, \alpha_2, \beta_2, \alpha_3, \beta_3, \gamma_3)$.\\
\texttt{\detokenize{t_min}} & minimum value for time.\\
\texttt{\detokenize{t_max}} & maximum value for time.\\
\texttt{\detokenize{anchor_point}} & (optional) vector of \texttt{\detokenize{(x, y)}} coordinates of the point to condition on. If \texttt{\detokenize{NULL}}, inferred from the reference data (largest mark if available) or from \texttt{\detokenize{process_fit$data_original}} (largest size).\\
\addlinespace
\texttt{\detokenize{raster_list}} & (optional) list of raster objects used for mark prediction. Required when \texttt{\detokenize{mark_mode = ``mark_model''}} unless rasters are stored in \texttt{\detokenize{mark_model}}.\\
\texttt{\detokenize{scaled_rasters}} & \texttt{\detokenize{TRUE}} or \texttt{\detokenize{FALSE}} indicating whether rasters are already scaled. Ignored when \texttt{\detokenize{mark_mode = ``time_to_size''}}.\\
\texttt{\detokenize{mark_model}} & (optional) mark model object used when \texttt{\detokenize{mark_mode = ``mark_model''}}. May be an \texttt{\detokenize{ldmppr_mark_model}}, \texttt{\detokenize{model_fit}}, or \texttt{\detokenize{workflow}}.\\
\texttt{\detokenize{xy_bounds}} & (optional) vector of bounds as \texttt{\detokenize{c(a_x, b_x, a_y, b_y)}}. If \texttt{\detokenize{NULL}}, bounds are inferred from \texttt{\detokenize{process_fit}} when available.\\
\texttt{\detokenize{include_comp_inds}} & \texttt{\detokenize{TRUE}} or \texttt{\detokenize{FALSE}} indicating whether to compute competition indices.\\
\addlinespace
\texttt{\detokenize{competition_radius}} & distance for competition radius if \texttt{\detokenize{include_comp_inds = TRUE}}.\\
\texttt{\detokenize{thinning}} & \texttt{\detokenize{TRUE}} or \texttt{\detokenize{FALSE}} indicating whether to use the thinned simulated values.\\
\texttt{\detokenize{edge_correction}} & type of edge correction to apply (\texttt{\detokenize{``none''}} or \texttt{\detokenize{``toroidal''}}).\\
\texttt{\detokenize{seed}} & integer seed for reproducibility.\\
\texttt{\detokenize{mark_mode}} & (optional) mark generation mode. \texttt{\detokenize{``mark_model''}} uses \texttt{\detokenize{predict()}} on a mark model, while \texttt{\detokenize{``time_to_size''}} maps simulated times back to sizes via \texttt{\detokenize{delta}}. If \texttt{\detokenize{NULL}}, inferred as \texttt{\detokenize{``mark_model''}} when \texttt{\detokenize{mark_model}} is provided, otherwise \texttt{\detokenize{``time_to_size''}}.\\
\addlinespace
\texttt{\detokenize{size_range}} & (optional) numeric vector \texttt{\detokenize{c(smin, smax)}} used for \texttt{\detokenize{mark_mode = ``time_to_size''}}. If \texttt{\detokenize{NULL}}, inferred from \texttt{\detokenize{process_fit}} when possible.\\
\texttt{\detokenize{delta}} & (optional) positive scalar used for \texttt{\detokenize{mark_mode = ``time_to_size''}}. If \texttt{\detokenize{NULL}}, inferred from \texttt{\detokenize{process_fit}} when possible.\\
\bottomrule
\end{longtable}

\ltd{The function returns a \texttt{ldmppr\_sim} object containing the relevant generating information, which can be provided to the function directly or inferred by passing a \texttt{ldmppr\_fit} and \texttt{ldmppr\_mark\_model} through the \texttt{process\_fit} and \texttt{mark\_model} arguments respectively, and realized simulation in both a \texttt{ppp} format and as a \texttt{data.frame}. With a realization of the process in hand, the user
can easily visualize the marked point process object using the default \texttt{plot()} method, which has the following arguments.}

\renewcommand{\arraystretch}{1.1}
\begin{longtable}{>{\raggedleft\arraybackslash}p{0.26\linewidth}p{0.68\linewidth}}
\toprule
Argument & Description\\
\midrule
\texttt{\detokenize{x}} & an \texttt{\detokenize{ldmppr_sim}} object\\
\texttt{\detokenize{pattern_type}} & type of pattern to plot \texttt{\detokenize{``simulated''}} (default).\\
\bottomrule
\end{longtable}

In the following section, we walk through the entire workflow described
in Section \hyperref[workflow]{3.1} with an example forestry dataset.

\section{Application}\label{sec:application}

Equipped with an understanding of the package workflow and primary
functionality, we now provide an example of using \CRANpkg{ldmppr} to
analyze a forest stand dataset comprised of canopy volumes (in cubic
meters) and locations for conifer species in the Southern Rocky
Mountains, obtained from \citet{drewDataBayesianRecord2024}, that is included
in the package. We incorporate four topographic covariates, in the form
of raster surfaces, that have been previously found to be related to the
processes of tree growth and that capture key environmental conditions
like energy and water availability \citep{drewBayesianRecordLinkage2025}. The
covariates included in this analysis are Southness Aspect, Topographic
Wetness Index, Elevation, and Slope which are derived from a LiDAR based
digital elevation model (DEM).

Following the steps outlined in Section \hyperref[sec:package]{3}, we begin by
estimating the self-correcting model using \(\delta=1\) as the size to
time mapping parameter. We load the data and specify the mapping as follows.

\begin{verbatim}
data("medium_example_data")
parameter_estimation_data <- medium_example_data 

delta <- 1
\end{verbatim}

\ltd{Next, we define the grid values for the spatial and temporal components
of the process and upper bounds using the \texttt{ldmppr\_grids()} function, and the budgets for the optimization procedure using the \texttt{ldmppr\_budgets()} function. To get a sense of where to start with the grid values, we recommend plotting the data and considering the spatial and temporal scales of the observed dynamics. For example, in this dataset, the spatial domain is approximately \(50\)m by \(50\)m, and the mapped arrival times range from 0 to 1. We want to ensure that the grid values capture the relevant scales of the observed dynamics, while balancing computational cost. The nearest neighbor distance distribution in our dataset can be used to identify a reasonable starting point for the spatial grid values.}

\begin{verbatim}
summary(spatstat.geom::nndist(parameter_estimation_data[,-3]))
\end{verbatim}

\begin{verbatim}
#>    Min. 1st Qu.  Median    Mean 3rd Qu.    Max. 
#>  0.9415  2.3064  2.7598  2.9095  3.4448  6.4035
\end{verbatim}

\ltd{Given this distribution, we start with a single grid level, specified as a \(10 \times 10 \times 10\) grid for the spatial and temporal components, respectively. We opt to employ the \texttt{“global\_local”} strategy for optimization, which includes a global optimization stage followed by a local polish. We start with a moderate number of function evaluations and higher tolerance for both the global and local stages to facilitate a low cost initial estimation.}

\begin{verbatim}
upper_bounds <- c(1, 50, 50)

grids <- ldmppr_grids(
  upper_bounds = upper_bounds,
  levels = list(c(10, 10, 10))
)

budgets <- ldmppr_budgets(
  global_options = list(
    maxeval = 500,
    ftol_rel = 1e-3,
    xtol_rel = 1e-3
  ),
  local_budget_first_level = list(
    maxeval = 1000,
    ftol_rel = 1e-4,
    xtol_rel = 1e-4
  )
)

starts <- list(global = 1, local = 1, jitter_sd = 0.1, seed = 90210)
\end{verbatim}

\ltd{Using the \texttt{estimate\_process\_parameters()} function, we then estimate the parameters of the self-correcting process and obtain the optimal parameter estimates. We use the \texttt{NLOPT\_GN\_CRS2\_LM} global and \texttt{NLOPT\_LN\_BOBYQA} local algorithms (described in Section \hyperref[scestimation]{3.2}) with the internally derived initial values and parameter bounds based on the reference data for the optimization. Details regarding the derivation of the initial values and bounds can be found in Appendix A of the Supplementary Materials.}

\begin{verbatim}
estimated_sc <- estimate_process_parameters(
  data = parameter_estimation_data,
  process = "self_correcting",
  grids = grids,
  budgets = budgets,
  delta = 1,
  parallel = FALSE,
  strategy = "global_local",
  global_algorithm = "NLOPT_GN_CRS2_LM",
  local_algorithm = "NLOPT_LN_BOBYQA",
  starts = starts,
  verbose = FALSE
)

optimal_parameters <- coef(estimated_sc)
summary(estimated_sc)
\end{verbatim}

\begin{verbatim}
#> Summary: ldmppr Fit
#>   process:         self_correcting
#>   engine:          nloptr
#>   strategy:        global_local
#>   starts:          global=1, local=1, jitter_sd=0.35, seed=90210
#>   status:          3
#>   outcome:         NLOPT_FTOL_REACHED: Optimization stopped because ftol_rel or
#>                    ftol_abs (above) was reached.
#>   objective:       432.2978
#>   selected_delta:  1
#>   elapsed_sec:     0.038
#>   coefficients:
#> [1] 4.516536e+00 2.495385e-01 8.525285e-08 1.129760e-07 1.105553e-07
#> [6] 1.095338e-07 1.067978e-07 1.205356e-01
\end{verbatim}

We obtain the estimated parameter set
\((\alpha_1 = 4.5165\),
\(\beta_1 = 0.2495\),
\(\gamma_1 = 0\),
\(\alpha_2 = 0\),
\(\beta_2 = 0\),
\(\alpha_3 = 0\),
\(\beta_3 = 0\),
\(\gamma_3 = 0.1205)\). In practice, we
recommend starting with a low budget initial fit to gauge the need for increasing the number of iterations and reducing the
fractional tolerance levels for the optimization procedure, and then adjusting these settings as needed to improve the fit.

In addition to estimating the self-correcting process, we need to train
the conditional mark model for canopy volume (\(m^3\)) given the arrival
times, locations, location specific topographic covariates, and
interpoint competition metrics in a \(10m\) neighborhood. We begin by
loading the example raster images and scaling them.

\begin{verbatim}
raster_paths <- list.files(system.file("extdata", package = "ldmppr"),
                           pattern = "\\.tif$", full.names = TRUE)
raster_paths <- raster_paths[grepl("_med\\.tif$", raster_paths)]
rasters <- lapply(raster_paths, terra::rast)
scaled_rasters <- scale_rasters(rasters)
\end{verbatim}

Next, to train the mark model, \ltd{we opt for a gradient boosted tree model
using the \CRANpkg{xgboost} engine with 5-fold cross-validation and a hyperparameter tuning grid of size 50. We can pass the prepared data directly from the previous step, and specify the use of parallelization to speed up the training process. We use the selection metric of mean absolute error (MAE) to identify the optimal model from the tuning grid.}

\begin{verbatim}
example_trained_mark_model <- train_mark_model(
  data = estimated_sc,
  raster_list = scaled_rasters,
  scaled_rasters = TRUE,
  model_type = "xgboost",
  parallel = TRUE,
  include_comp_inds = TRUE,
  competition_radius = 10,
  edge_correction = "none",
  selection_metric = "mae",
  cv_folds = 5,
  tuning_grid_size = 50,
  verbose = FALSE
)
\end{verbatim}

With the estimated parameters for the self-correcting process and a
\ltd{trained conditional mark model in hand, we now assess how well the estimated process reflects the dynamics in
the original dataset using the \texttt{check\_model\_fit()} function. In order
to obtain valid \(p\)-values at the \(\alpha = .05\) level for the global
envelope tests, \citet{myllymakiGlobalEnvelopeTests2017} recommend using a
minimum of 2499 realizations from the process, however for our initial fit we reduce the number of realizations to get an approximate sense of our performance. We pass the reference data and estimated process parameters from our estimation step through the \texttt{process\_fit} argument, and the trained mark model and scaled rasters through the \texttt{mark\_model} argument. We also include the competition indices in the simulations to match the training of the mark model.}

\begin{verbatim}
example_model_fit <- check_model_fit(
  process = "self_correcting",
  process_fit = estimated_sc,
  mark_model = example_trained_mark_model,
  include_comp_inds = TRUE,
  thinning = TRUE,
  edge_correction = "none",
  competition_radius = 10,
  n_sim = 499,
  save_sims = FALSE,
  verbose = FALSE,
  parallel = FALSE,
  fg_correction = "km"
)
\end{verbatim}

Once we have run the \texttt{check\_model\_fit()} function, we can plot the
combined global envelope test, or any of the individual tests for the
summary functions (\(L\), \(F\), \(G\), \(J\), \(E\), or \(V\)), using the default \texttt{plot()} method as seen in Figure
\ref{fig:latexenvplot}.

\begin{verbatim}
plot(example_model_fit)
\end{verbatim}

In this example, our initial estimated model does not actually provide evidence of an
adequate fit to the reference dataset as evidenced by the small \(p\)
value (\(p\) = 0.002) and the number of points from the reference data that fall outside
of the simulated envelopes (as highlighted in red). \ltd{In particular, the
nearest neighbor distance distribution function, \(G(r)\), from the original data is not
well captured by realizations from the fitted process. The \(G(r)\)
function is the distribution function of the distance from an arbitrary event to its nearest neighboring event, and is commonly used to assess the degree of clustering or inhibition in a point pattern relative to complete spatial randomness (CSR)
\citep{ripleyStatisticalInferenceSpatial1988}}. We also note that the function
\(V(r)\), which captures the variance of the mark associated with a
typical random point given that another random point exists at distance
\(r\), is not well captured by the fitted process. This suggests that the
estimated mark model may not be capturing the dynamics of the mark
process well, or that the self-correcting process is not adequately
capturing the spatial dynamics of the reference dataset.

\begin{figure}
\centering
\pandocbounded{\includegraphics[keepaspectratio,alt={\label{fig:latexenvplot}Combined global envelope test for the realizations from the fitted process. Solid black lines represent the reference process, dashed black lines represent a homogeneous Poisson process (or CSR), and the colored band represents the global envelope for the simulated datasets at the \textbackslash alpha = .05 level. Reference values outside the envelope are highlighted in red and suggest a poor fit.}]{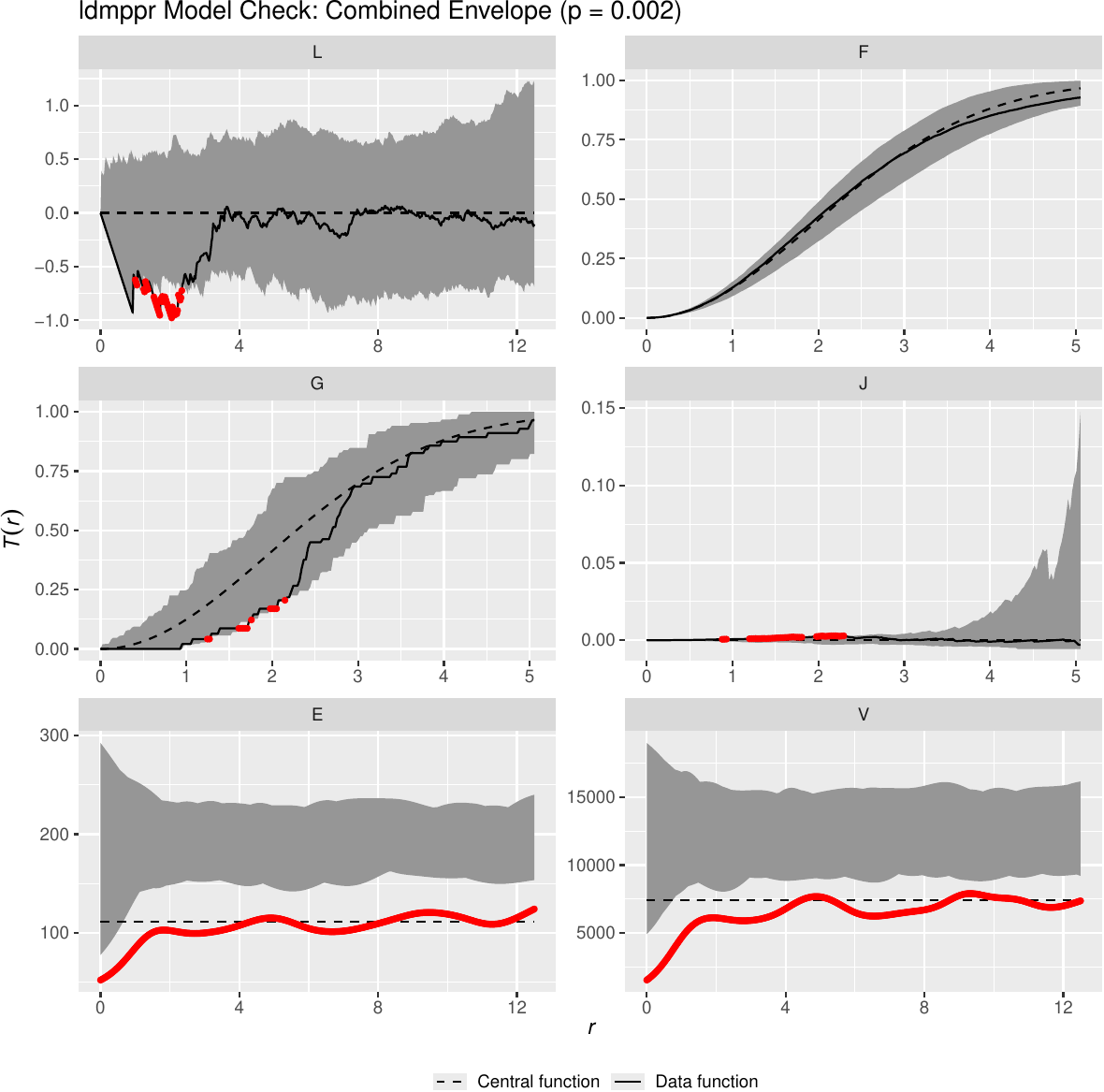}}
\caption{\label{fig:latexenvplot}Combined global envelope test for the realizations from the fitted process. Solid black lines represent the reference process, dashed black lines represent a homogeneous Poisson process (or CSR), and the colored band represents the global envelope for the simulated datasets at the \(\alpha = .05\) level. Reference values outside the envelope are highlighted in red and suggest a poor fit.}
\end{figure}

This result suggests that we may need to increase the budget and strategy for our self-correcting model estimation step when using
the \texttt{estimate\_process\_parameters()} function, and potentially expand the training regime for our conditional mark model.

\ltd{Given that we obtained a less than optimal fit with our initial modeling
attempt, we revisit both the self-correcting process and mark models to
see if we can improve them. A standard approach for improving estimation
of the self-correcting process is to increase the granularity of the
grid used to estimate the model, while increasing the number of
iterations and potentially reducing the tolerance threshold. We can also leverage the \texttt{“multires\_global\_local”} strategy to improve the estimate iteratively, while making use of multiple starts at each level to expand our search area to identify a global optimum. We refit the model with a sequence of finer grid levels and increase the budget for the optimization procedure to allow for a more thorough search of the parameter space, and reduce the tolerance levels to facilitate convergence to a more polished solution. We also introduce additional starts at each optimization level to explore the parameter space more effectively and reduce our dependence on the initial values. We maintain the same optimization strategy and algorithms as before, and use the internally derived initial values and parameter bounds based on the reference data for the optimization.}

\begin{verbatim}
grids <- ldmppr_grids(
  upper_bounds = upper_bounds,
  levels = list(c(10, 10, 10),
                c(12, 12, 12),
                c(16, 16, 16))
)

budgets <- ldmppr_budgets(
  global_options = list(
    maxeval = 1000,
    ftol_rel = 1e-6,
    xtol_rel = 1e-6
  ),
  local_budget_first_level = list(
    maxeval = 1500,
    ftol_rel = 1e-8,
    xtol_rel = 1e-8
  ),
  local_budget_refinement_levels = list(
    maxeval = 2000,
    ftol_rel = 1e-10,
    xtol_rel = 1e-10
  )
  
)

estimated_sc_update <- estimate_process_parameters(
  data = parameter_estimation_data,
  process = "self_correcting",
  grids = grids,
  budgets = budgets,
  delta = 1,
  parallel = FALSE,
  strategy = "multires_global_local",
  global_algorithm = "NLOPT_GN_CRS2_LM",
  local_algorithm = "NLOPT_LN_BOBYQA",
  starts = list(global = 5, local = 3, jitter_sd = 0.15, seed = 90210),
  verbose = TRUE
)

improved_optimal_parameters <- coef(estimated_sc_update)
summary(estimated_sc_update)
\end{verbatim}

We obtain the following improved parameter estimates from the
optimization.

\begin{verbatim}
improved_optimal_parameters <- if (inherits(estimated_sc_update, "ldmppr_fit")) {
  coef(estimated_sc_update)
} else {
  as.numeric(estimated_sc_update$solution)
}
summary(estimated_sc_update)
\end{verbatim}

\begin{verbatim}
#> Summary: ldmppr Fit
#>   process:         self_correcting
#>   engine:          nloptr
#>   strategy:        multires_global_local
#>   starts:          global=5, local=3, jitter_sd=0.15, seed=90210
#>   status:          3
#>   outcome:         NLOPT_FTOL_REACHED: Optimization stopped because ftol_rel or
#>                    ftol_abs (above) was reached.
#>   objective:       387.0972
#>   selected_delta:  1
#>   elapsed_sec:     1.935
#>   coefficients:
#> [1] 1.4405824 7.3756717 0.0314348 1.6980704 0.9817515 2.4976881 1.8469612
#> [8] 0.0832517
\end{verbatim}

We see that the improved parameter estimates set
\((\alpha_1 = 1.4406\),
\(\beta_1 = 7.3757\),
\(\gamma_1 = 0.0314\),
\(\alpha_2 = 1.6981\),
\(\beta_2 = 0.9818\),
\(\alpha_3 = 2.4977\),
\(\beta_3 = 1.847\),
\(\gamma_3 = 0.0833)\) has shifted
compared to the initial estimates.

Next, we retrain the mark model using two changes. \ltd{First, we increase the size of the hyperparameter tuning grid to 150 to better explore the
hyperparameter space and increase the number of cross-validation folds from 5 to 10, which typically reduces the variance of performance
estimates to provide a more stable estimate of model generalization.}

We then retrain the mark model using the updated estimated process and adjusted model specification.

\begin{verbatim}
improved_example_trained_mark_model <- train_mark_model(
  data = estimated_sc_update,
  xy_bounds = c(0, 50, 0, 50),
  raster_list = scaled_rasters,
  scaled_rasters = TRUE,
  model_type = "xgboost",
  parallel = TRUE,
  include_comp_inds = TRUE,
  competition_radius = 10,
  edge_correction = "none",
  selection_metric = "mae",
  cv_folds = 10,
  tuning_grid_size = 150,
  verbose = TRUE
)
\end{verbatim}

We proceed by checking the fit of our model using the improved parameter
estimates for the self-correcting process and the retrained mark model.
We use the same reference dataset and raster images as before, and
simulate a full 2500 realizations from the process to assess the fit of the
updated model.

\begin{verbatim}
improved_example_model_fit <- check_model_fit(
  process = "self_correcting",
  process_fit = estimated_sc_update,
  mark_model = improved_example_trained_mark_model,
  include_comp_inds = TRUE,
  thinning = TRUE,
  edge_correction = "none",
  competition_radius = 10,
  n_sim = 2500,
  save_sims = FALSE,
  verbose = TRUE,
  parallel = FALSE,
  fg_correction = "km"
)
\end{verbatim}

Next, we check the combined global envelope test for the updated model
to assess how well the estimated model captures the dynamics of the
reference dataset.

\begin{verbatim}
plot(improved_example_model_fit)
\end{verbatim}

\begin{figure}
\centering
\pandocbounded{\includegraphics[keepaspectratio,alt={\label{fig:latexenvplot2}Combined global envelope test for the realizations from the improved fitted process. As in Figure , solid black lines represent the reference process, dashed black lines represent a homogeneous Poisson process (or CSR), and the colored band represents the global envelope for the simulated datasets at the \textbackslash alpha = .05 level.}]{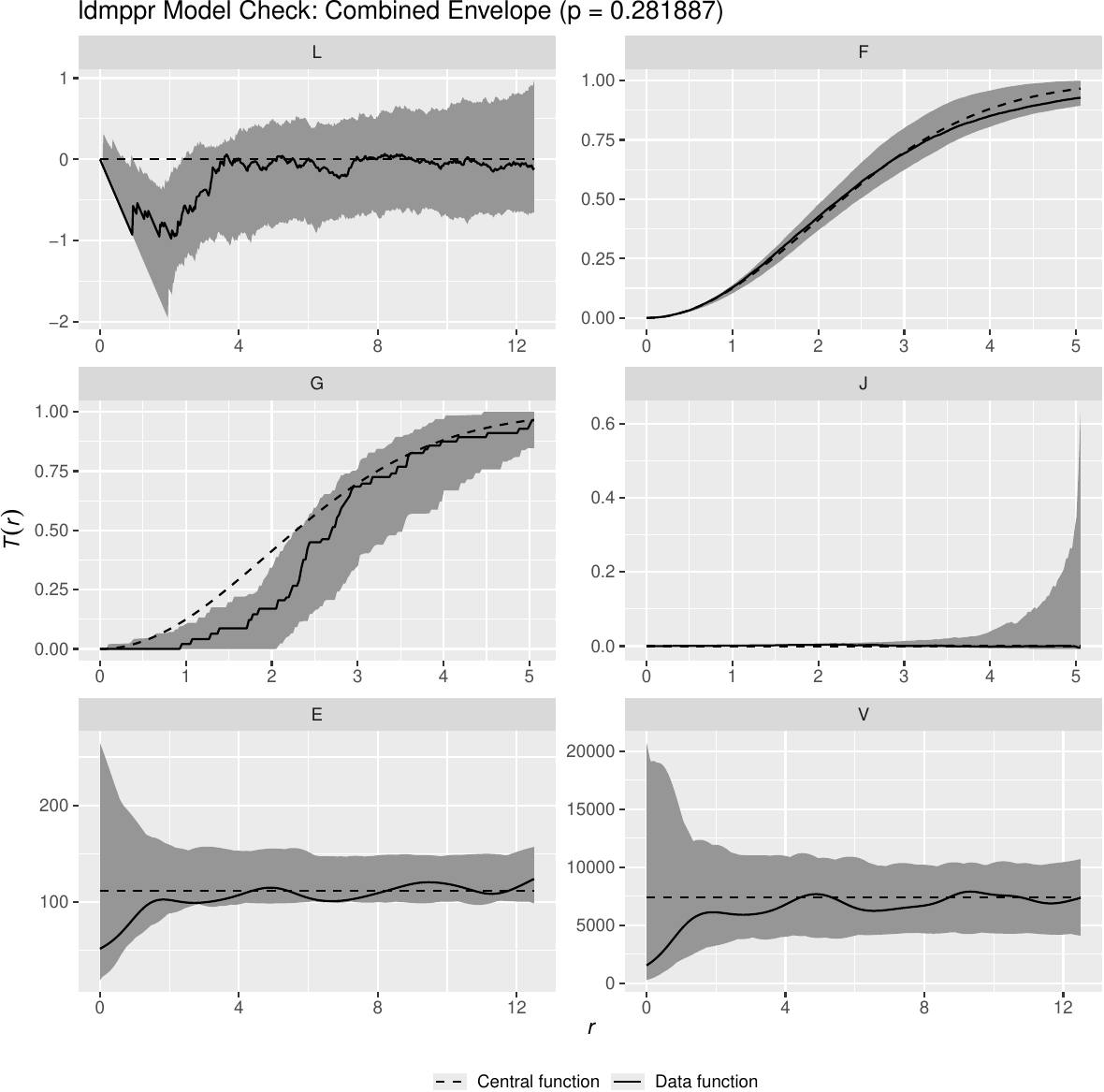}}
\caption{\label{fig:latexenvplot2}Combined global envelope test for the realizations from the improved fitted process. As in Figure \ref{fig:latexenvplot}, solid black lines represent the reference process, dashed black lines represent a homogeneous Poisson process (or CSR), and the colored band represents the global envelope for the simulated datasets at the \(\alpha = .05\) level.}
\end{figure}

We see that the improved model provides a notably improved fit to the reference
dataset, as seen in Figure \ref{fig:latexenvplot2}, evidenced by the fact that the simulation
envelopes contain the reference pattern across all six metrics. In
addition, the \(p\)-value for the combined global envelope test is
\(p = 0.2819\),
indicating that the estimated process is a markedly improved fit to the
reference dataset compared to the fit that we initially achieved.

When evaluating the fit of a model, in addition to the non-parametric
summary statistics and global envelope tests, it may also be useful to
perform a visual comparison of a realization from the model and the
reference dataset. To assess the agreement between the improved fitted
model and the reference data, we simulate a realization from the
improved model and compare it to the original reference dataset and a
dataset simulated from the initial (poorly fitting) model.

\begin{verbatim}
improved_simulated_mpp <- simulate_mpp(
  process = "self_correcting",
  process_fit = estimated_sc_update,
  mark_model = improved_example_trained_mark_model,
  include_comp_inds = TRUE,
  competition_radius = 10,
  edge_correction = "none",
  thinning = TRUE
)

initial_simulated_mpp <- simulate_mpp(
  process = "self_correcting",
  process_fit = estimated_sc,
  mark_model = example_trained_mark_model,
  include_comp_inds = TRUE,
  competition_radius = 10,
  edge_correction = "none",
  thinning = TRUE
)

ref_plot <- plot_mpp(
  mpp_data = reference_data,
  pattern_type = "reference"
)

improved_sim_plot <- plot(
  mpp_data = improved_simulated_mpp,
  pattern_type = "simulated"
)

initial_sim_plot <- plot(
  mpp_data = initial_simulated_mpp,
  pattern_type = "simulated"
)
\end{verbatim}

\begin{figure}
\centering
\pandocbounded{\includegraphics[keepaspectratio,alt={\label{fig:latexdataplot2}Plots a) - c) provide a comparison of a realization from the improved estimated process with the original reference dataset and a realization from the initial estimated process. Plot d) shows the corresponding observed mark distributions from the reference dataset and the simulation realizations.}]{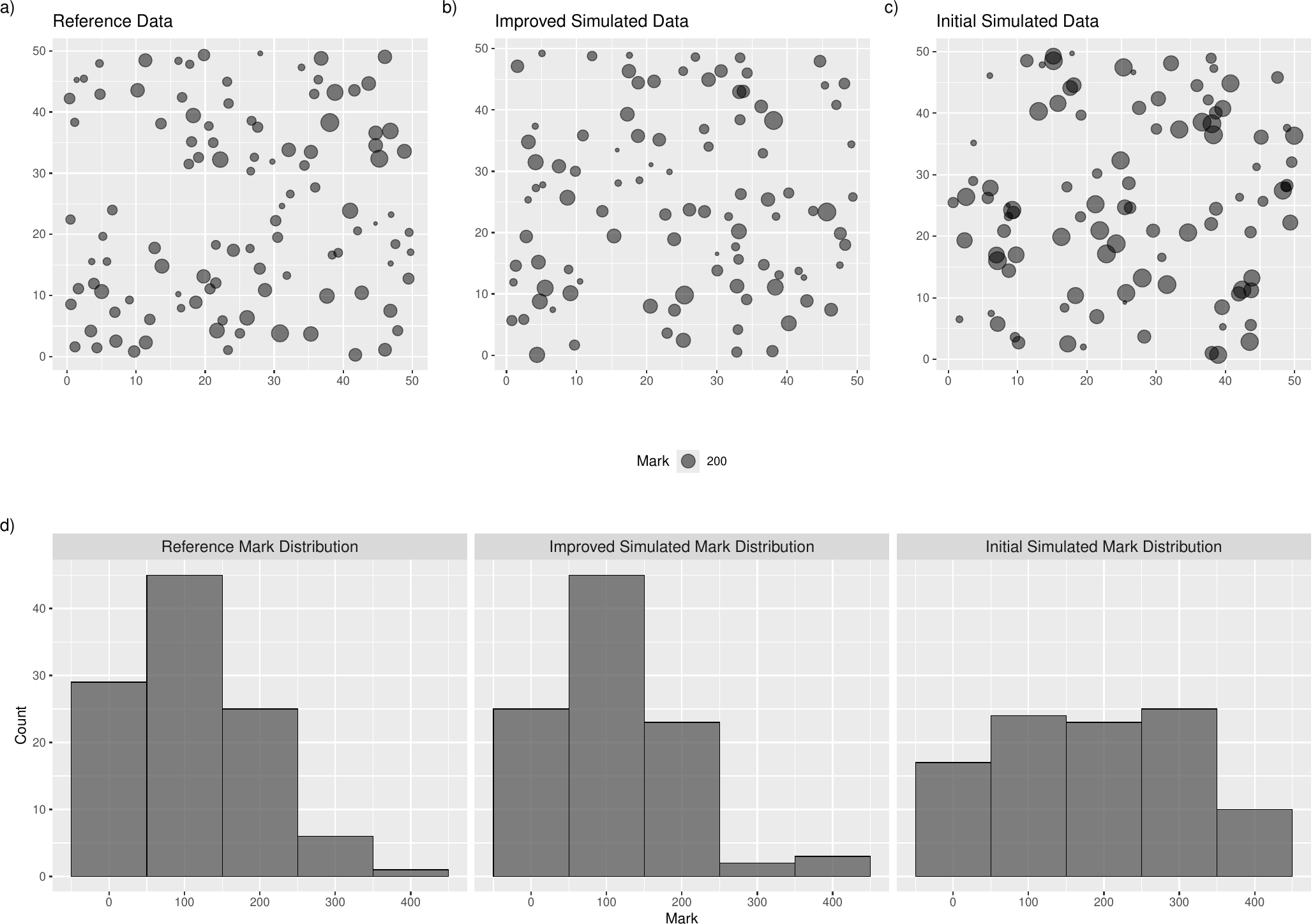}}
\caption{\label{fig:latexdataplot2}Plots a) - c) provide a comparison of a realization from the improved estimated process with the original reference dataset and a realization from the initial estimated process. Plot d) shows the corresponding observed mark distributions from the reference dataset and the simulation realizations.}
\end{figure}

Figure \ref{fig:latexdataplot2} demonstrates that the improved model
provides a more accurate representation of the reference dataset than we
obtained from the initial model in terms of the spatial process and the
conditional mark process. We also include histograms of the realized
mark distributions to highlight the improvement in replicating the mark
distribution of the reference process. This visualization provides
additional evidence that the improved model captures the spatial and
mark dynamics of the reference dataset more effectively, which results
in the improved model being a better fit. This example highlights the
utility of the package for working with a real marked point process of
interest, and provides the intuition for using the main functionality of
\CRANpkg{ldmppr} to estimate, assess model fit, and simulate from a marked
spatial point process.

\ltd{We note that the computational cost of the workflow is driven primarily by three steps: self-correcting process estimation, mark-model training with cross-validation/tuning, and goodness-of-fit simulation in \texttt{check\_model\_fit()}. Runtime increases with grid resolution, optimizer budgets, number of starts, tuning grid size, cross-validation folds, and the number of simulations (\texttt{n\_sim}). A practical approach is to begin with a low-cost baseline (coarser grid, fewer starts, smaller tuning grid, moderate \texttt{n\_sim}) and increase complexity only when diagnostics indicate the need for refinement. Parallelization can also reduce wall-clock time for multi-start estimation, model tuning, and simulation-based checks when working with larger datasets, though the benefits are less pronounced for small datasets or lower budget runs. Concrete benchmark tables are provided in Appendices B and C of the Supplementary Materials for various usage scenarios, and Appendix D provides a contextual side-by-side comparison against a two-stage \pkg{spatstat} derived workflow under matched tuning/check settings.}

\section{Discussion}\label{sec:discussion}

In this paper, we provide a novel framework for estimating location
dependent marked point processes and introduce the \CRANpkg{ldmppr} package,
which contains a user-friendly modular suite of tools for model
estimation, evaluation, simulation, and visualization for marked point
processes with location dependence characterized by regularity in the
spatial pattern. We outlined the typical workflow for using the package
and discussed the key functions and their arguments in detail before
providing an example of using the package with a real forestry dataset.
\CRANpkg{ldmppr} simplifies the process of working with marked point
processes and provides a likelihood based estimation approach that is
computationally feasible. While the framework presented applies to a
broader range of marked point process, the package in its current
implementation is still somewhat limited in the types of patterns that
it can address. As the package continues to develop, we would like to incorporate additional models that can
address point process data that demonstrates clustering behavior, as
opposed to regularity, while still maintaining the focus on location
dependent marks.

\bibliography{RJrefs.bib}

\clearpage
\section*{Supplementary Materials for ldmppr: Location Dependent Marked Point Processes in R}
\appendix

\section{Additional details on initialization defaults for \\ \texttt{\detokenize{estimate_process_parameters}()}}\label{appA}

We initialize self-correcting model parameters using data-adaptive
heuristics designed to place the optimizer in a stable region of the
parameter space. The baseline log-intensity is anchored by the empirical
event rate via \(\alpha_{\text{base}} = \log(n / \Delta t)\). The
self-correction strength \(\gamma_1\) is initialized on a conservative
\(\log(n) / n\)-type scale to reduce unstable behavior early in
optimization. To capture monotone temporal trends, \(\beta_1\) is
initialized from a Poisson regression of binned event counts on time
(with an offset for bin width), with a fallback that induces a modest
multiplicative change across the observation window.

Spatial inhibition range \(\alpha_2\) is initialized using the median
nearest-neighbor distance (or a conservative geometric fallback).
Spatio-temporal interaction scales are bounded by the observation
window: the spatial scale parameter is capped by the window diagonal,
and the temporal interaction scale is capped by the observed time span
(equal to 1 under the default time mapping). These choices provide
stable starting values and natural finite bounds for optimization.

\section{Stage-wise runtime benchmarks}\label{appB}

To quantify fitting cost, we benchmarked runtime for
\texttt{\detokenize{estimate_process_parameters}()} and
\texttt{\detokenize{train_mark_model}()} on three contiguous datasets:
(i) \texttt{medium\_example\_data}, (ii) a contiguous window with
approximately 200 points, and (iii) a contiguous window with
approximately 400 points. The larger datasets were selected as
contiguous windows from the full domain (rather than random point
subsamples) to preserve local dependence structure.

Stage-wise timing artifacts can be regenerated by running: \newline
\texttt{LDMPPR\_STAGE\_PROFILE=full} \newline
\texttt{LDMPPR\_STAGE\_REPS=10} \newline
\texttt{LDMPPR\_STAGE\_CORES=1,7} \newline
\texttt{LDMPPR\_STAGE\_DATASETS=medium,win200,win400} \newline
\texttt{Rscript "scripts/supplement\_stage\_timing\_study.R"}.

If \texttt{LDMPPR\_RASTER\_DIR} is not set, required external rasters
are downloaded automatically from ESS-DIVE into
\texttt{data/ess\_dive\_rasters}.

\begin{longtable}[]{@{}
  >{\raggedright\arraybackslash}p{(\linewidth - 20\tabcolsep) * \real{0.0833}}
  >{\raggedleft\arraybackslash}p{(\linewidth - 20\tabcolsep) * \real{0.0729}}
  >{\raggedleft\arraybackslash}p{(\linewidth - 20\tabcolsep) * \real{0.1250}}
  >{\raggedleft\arraybackslash}p{(\linewidth - 20\tabcolsep) * \real{0.0625}}
  >{\raggedleft\arraybackslash}p{(\linewidth - 20\tabcolsep) * \real{0.0521}}
  >{\raggedleft\arraybackslash}p{(\linewidth - 20\tabcolsep) * \real{0.1042}}
  >{\raggedleft\arraybackslash}p{(\linewidth - 20\tabcolsep) * \real{0.0833}}
  >{\raggedleft\arraybackslash}p{(\linewidth - 20\tabcolsep) * \real{0.1146}}
  >{\raggedleft\arraybackslash}p{(\linewidth - 20\tabcolsep) * \real{0.0938}}
  >{\raggedleft\arraybackslash}p{(\linewidth - 20\tabcolsep) * \real{0.1146}}
  >{\raggedleft\arraybackslash}p{(\linewidth - 20\tabcolsep) * \real{0.0938}}@{}}
\caption{Stage-wise runtime summary (seconds): process estimation and
mark-model training.}\tabularnewline
\toprule\noalign{}
\begin{minipage}[b]{\linewidth}\raggedright
Dataset
\end{minipage} & \begin{minipage}[b]{\linewidth}\raggedleft
Points
\end{minipage} & \begin{minipage}[b]{\linewidth}\raggedleft
Window Side
\end{minipage} & \begin{minipage}[b]{\linewidth}\raggedleft
Cores
\end{minipage} & \begin{minipage}[b]{\linewidth}\raggedleft
Runs
\end{minipage} & \begin{minipage}[b]{\linewidth}\raggedleft
Est. Mean
\end{minipage} & \begin{minipage}[b]{\linewidth}\raggedleft
Est. SD
\end{minipage} & \begin{minipage}[b]{\linewidth}\raggedleft
Train Mean
\end{minipage} & \begin{minipage}[b]{\linewidth}\raggedleft
Train SD
\end{minipage} & \begin{minipage}[b]{\linewidth}\raggedleft
Total Mean
\end{minipage} & \begin{minipage}[b]{\linewidth}\raggedleft
Total SD
\end{minipage} \\
\midrule\noalign{}
\endfirsthead
\toprule\noalign{}
\begin{minipage}[b]{\linewidth}\raggedright
Dataset
\end{minipage} & \begin{minipage}[b]{\linewidth}\raggedleft
Points
\end{minipage} & \begin{minipage}[b]{\linewidth}\raggedleft
Window Side
\end{minipage} & \begin{minipage}[b]{\linewidth}\raggedleft
Cores
\end{minipage} & \begin{minipage}[b]{\linewidth}\raggedleft
Runs
\end{minipage} & \begin{minipage}[b]{\linewidth}\raggedleft
Est. Mean
\end{minipage} & \begin{minipage}[b]{\linewidth}\raggedleft
Est. SD
\end{minipage} & \begin{minipage}[b]{\linewidth}\raggedleft
Train Mean
\end{minipage} & \begin{minipage}[b]{\linewidth}\raggedleft
Train SD
\end{minipage} & \begin{minipage}[b]{\linewidth}\raggedleft
Total Mean
\end{minipage} & \begin{minipage}[b]{\linewidth}\raggedleft
Total SD
\end{minipage} \\
\midrule\noalign{}
\endhead
\bottomrule\noalign{}
\endlastfoot
medium & 106 & 50 & 1 & 10 & 2.91 & 0.33 & 23.34 & 0.76 & 26.25 &
0.74 \\
medium & 106 & 50 & 7 & 10 & 2.23 & 0.27 & 23.76 & 0.15 & 25.99 &
0.28 \\
win200 & 207 & 75 & 1 & 10 & 7.72 & 1.09 & 36.05 & 0.16 & 43.76 &
1.14 \\
win200 & 207 & 75 & 7 & 10 & 4.77 & 0.79 & 35.45 & 0.19 & 40.22 &
0.72 \\
win400 & 413 & 100 & 1 & 10 & 19.95 & 1.87 & 60.20 & 0.52 & 80.15 &
1.82 \\
win400 & 413 & 100 & 7 & 10 & 7.74 & 2.02 & 60.08 & 0.29 & 67.82 &
2.06 \\
\end{longtable}

Using one core, process-estimation time increases from 2.91s (n=106) to
19.95s (n=413), while mark-model training increases from 23.34s to
60.20s over the same range.

\section{Simulation-check runtime benchmarks}\label{appC}

We benchmarked runtime for \texttt{\detokenize{check_model_fit}()} using
\texttt{n\_sim} values of 200, 500, 1000, and 2500, with fixed seeds and
matched settings otherwise. This isolates simulation-based
goodness-of-fit cost and provides practical guidance for exploratory
versus final checks.

Simulation-check timing artifacts can be regenerated by running:
\newline \texttt{LDMPPR\_TIMING\_MODE=full} \newline
\texttt{LDMPPR\_TIMING\_N\_SIM=200,500,1000,2500} \newline
\texttt{LDMPPR\_TIMING\_REPS=10} \newline
\texttt{LDMPPR\_TIMING\_CORES=1,2} \newline
\texttt{Rscript "scripts/supplement\_sim\_timing\_study.R"}.

\begin{longtable}[]{@{}
  >{\raggedleft\arraybackslash}p{(\linewidth - 10\tabcolsep) * \real{0.1690}}
  >{\raggedleft\arraybackslash}p{(\linewidth - 10\tabcolsep) * \real{0.0845}}
  >{\raggedleft\arraybackslash}p{(\linewidth - 10\tabcolsep) * \real{0.0704}}
  >{\raggedleft\arraybackslash}p{(\linewidth - 10\tabcolsep) * \real{0.1831}}
  >{\raggedleft\arraybackslash}p{(\linewidth - 10\tabcolsep) * \real{0.1549}}
  >{\raggedleft\arraybackslash}p{(\linewidth - 10\tabcolsep) * \real{0.3380}}@{}}
\caption{Runtime summary (seconds) for simulation-based model
checking.}\tabularnewline
\toprule\noalign{}
\begin{minipage}[b]{\linewidth}\raggedleft
Simulations
\end{minipage} & \begin{minipage}[b]{\linewidth}\raggedleft
Cores
\end{minipage} & \begin{minipage}[b]{\linewidth}\raggedleft
Runs
\end{minipage} & \begin{minipage}[b]{\linewidth}\raggedleft
Elapsed Mean
\end{minipage} & \begin{minipage}[b]{\linewidth}\raggedleft
Elapsed SD
\end{minipage} & \begin{minipage}[b]{\linewidth}\raggedleft
Combined \(p\)-value Mean
\end{minipage} \\
\midrule\noalign{}
\endfirsthead
\toprule\noalign{}
\begin{minipage}[b]{\linewidth}\raggedleft
Simulations
\end{minipage} & \begin{minipage}[b]{\linewidth}\raggedleft
Cores
\end{minipage} & \begin{minipage}[b]{\linewidth}\raggedleft
Runs
\end{minipage} & \begin{minipage}[b]{\linewidth}\raggedleft
Elapsed Mean
\end{minipage} & \begin{minipage}[b]{\linewidth}\raggedleft
Elapsed SD
\end{minipage} & \begin{minipage}[b]{\linewidth}\raggedleft
Combined \(p\)-value Mean
\end{minipage} \\
\midrule\noalign{}
\endhead
\bottomrule\noalign{}
\endlastfoot
200 & 1 & 10 & 4.33 & 0.10 & 0.133 \\
200 & 2 & 10 & 4.08 & 0.17 & 0.136 \\
500 & 1 & 10 & 10.63 & 0.13 & 0.136 \\
500 & 2 & 10 & 9.43 & 0.19 & 0.146 \\
1000 & 1 & 10 & 21.79 & 0.16 & 0.215 \\
1000 & 2 & 10 & 19.37 & 0.72 & 0.151 \\
2500 & 1 & 10 & 52.74 & 0.35 & 0.227 \\
2500 & 2 & 10 & 49.21 & 0.44 & 0.217 \\
\end{longtable}

At one core, mean check time grows from 4.33s at n\_sim=200 to 52.74s at
n\_sim=2500.

For the largest run (n\_sim=2500), moving from one core to two cores
reduces mean wall-clock time by 6.7\%.

For iterative model development, moderate simulation counts can reduce
turnaround time. For final reporting, larger simulation counts provide
more stable Monte Carlo \(p\)-value estimates.

For a direct, reproducible benchmark of the improved paper workflow
(estimation + mark-model training + model check, matching the script in
the main paper), run: \newline \texttt{LDMPPR\_BENCH\_REPS=10} \newline
\texttt{Rscript "scripts/benchmark\_improved\_workflow\_timing.R"}.

\section{Description and interpretation of the \texttt{spatstat} comparison framework}\label{appD}

The comparison script \texttt{scripts/spatstat\_comp.R} implements a
two-stage baseline intended to approximate the improved manuscript
workflow as closely as possible while utilizing the \texttt{spatstat}
modeling framework:

\begin{enumerate}
\item A spatial point process is fit with \texttt{spatstat::ppm}, using raster-derived covariates and a Strauss interaction term for regularity/inhibition.
\item A separate XGBoost mark model is fit from location/raster features with optional competition-index features. To align with the improved paper workflow, tuning uses MAE with 10-fold cross-validation and a tuning grid size of 150.
\item Simulated spatial patterns from the fitted \texttt{ppm} model are marked using the fitted XGBoost model then evaluated with the same LGFJEV and combined GET-rank envelope machinery used in the package-facing checks.
\end{enumerate}

For the direct comparison below we use \texttt{n\_sim = 2500} for both
methods, matching the improved-paper model-check setting. For the
\texttt{ldmppr} workflow, we match the paper's specifications exactly:
estimation and model-check steps run without parallelization, while
mark-model training uses \texttt{parallel = TRUE}. Comparison artifacts
can be regenerated by running: \newline
\texttt{LDMPPR\_COMP\_N\_SIM=2500} \newline
\texttt{LDMPPR\_COMP\_PARALLEL=true} \newline
\texttt{Rscript "scripts/ldmppr\_comp\_timing.R"} and \newline
\texttt{LDMPPR\_SPATSTAT\_N\_SIM=2500} \newline
\texttt{LDMPPR\_SPATSTAT\_CV\_FOLDS=10} \newline
\texttt{LDMPPR\_SPATSTAT\_TUNING\_GRID=150} \newline
\texttt{LDMPPR\_SPATSTAT\_FG\_CORRECTION=km} \newline
\texttt{LDMPPR\_SPATSTAT\_PARALLEL=true} \newline
\texttt{LDMPPR\_SPATSTAT\_NUM\_CORES=7} \newline
\texttt{Rscript "scripts/spatstat\_comp.R"}.

If \texttt{LDMPPR\_RASTER\_DIR} is not set, required external rasters
are downloaded automatically from ESS-DIVE into
\texttt{data/ess\_dive\_rasters}.

\begin{longtable}[]{@{}
  >{\raggedright\arraybackslash}p{(\linewidth - 10\tabcolsep) * \real{0.2603}}
  >{\raggedleft\arraybackslash}p{(\linewidth - 10\tabcolsep) * \real{0.1644}}
  >{\raggedleft\arraybackslash}p{(\linewidth - 10\tabcolsep) * \real{0.1233}}
  >{\raggedleft\arraybackslash}p{(\linewidth - 10\tabcolsep) * \real{0.1370}}
  >{\raggedleft\arraybackslash}p{(\linewidth - 10\tabcolsep) * \real{0.1370}}
  >{\raggedleft\arraybackslash}p{(\linewidth - 10\tabcolsep) * \real{0.1781}}@{}}
\caption{Side-by-side timing and combined \(p\)-value comparison (ldmppr
vs spatstat two-stage) for \(n = 2500\) simulations in
seconds.}\tabularnewline
\toprule\noalign{}
\begin{minipage}[b]{\linewidth}\raggedright
Method
\end{minipage} & \begin{minipage}[b]{\linewidth}\raggedleft
Process fit
\end{minipage} & \begin{minipage}[b]{\linewidth}\raggedleft
Mark fit
\end{minipage} & \begin{minipage}[b]{\linewidth}\raggedleft
Check fit
\end{minipage} & \begin{minipage}[b]{\linewidth}\raggedleft
Total (s)
\end{minipage} & \begin{minipage}[b]{\linewidth}\raggedleft
Combined \(p\)
\end{minipage} \\
\midrule\noalign{}
\endfirsthead
\toprule\noalign{}
\begin{minipage}[b]{\linewidth}\raggedright
Method
\end{minipage} & \begin{minipage}[b]{\linewidth}\raggedleft
Process fit
\end{minipage} & \begin{minipage}[b]{\linewidth}\raggedleft
Mark fit
\end{minipage} & \begin{minipage}[b]{\linewidth}\raggedleft
Check fit
\end{minipage} & \begin{minipage}[b]{\linewidth}\raggedleft
Total (s)
\end{minipage} & \begin{minipage}[b]{\linewidth}\raggedleft
Combined \(p\)
\end{minipage} \\
\midrule\noalign{}
\endhead
\bottomrule\noalign{}
\endlastfoot
ldmppr & 7.49 & 136.54 & 53.84 & 197.87 & 0.2819 \\
spatstat two-stage & 0.02 & 128.59 & 443.81 & 443.81 & 0.0016 \\
\end{longtable}

Under matched comparison settings (n\_sim=2500), the combined GET
\(p\)-value is 0.2819 for ldmppr and 0.0016 for the two-stage spatstat
baseline.

Total runtime is 197.87s for ldmppr versus 443.81s for the two-stage
spatstat baseline; in the table above, the spatstat check time reports
simulation plus envelope-check computation, while the ldmppr check time
reports the corresponding integrated model-check step.

\begin{longtable}[]{@{}lrr@{}}
\caption{Side-by-side \(p\)-values by summary function, including the
combined test \(p\)-value.}\tabularnewline
\toprule\noalign{}
Statistic & ldmppr \(p\)-value & spatstat two-stage \(p\)-value \\
\midrule\noalign{}
\endfirsthead
\toprule\noalign{}
Statistic & ldmppr \(p\)-value & spatstat two-stage \(p\)-value \\
\midrule\noalign{}
\endhead
\bottomrule\noalign{}
\endlastfoot
L & 0.4650 & 0.0144 \\
F & 0.0728 & 0.5614 \\
G & 0.7137 & 0.1236 \\
J & 0.1443 & 0.0324 \\
E & 0.0688 & 0.0140 \\
V & 0.9100 & 0.0024 \\
Combined & 0.2819 & 0.0016 \\
\end{longtable}

At the 0.05 level, ldmppr shows significant departures for: none. The
two-stage spatstat baseline shows significant departures for: L, J, E,
V.

\begin{landscape}

\begin{center}\includegraphics[width=1\linewidth]{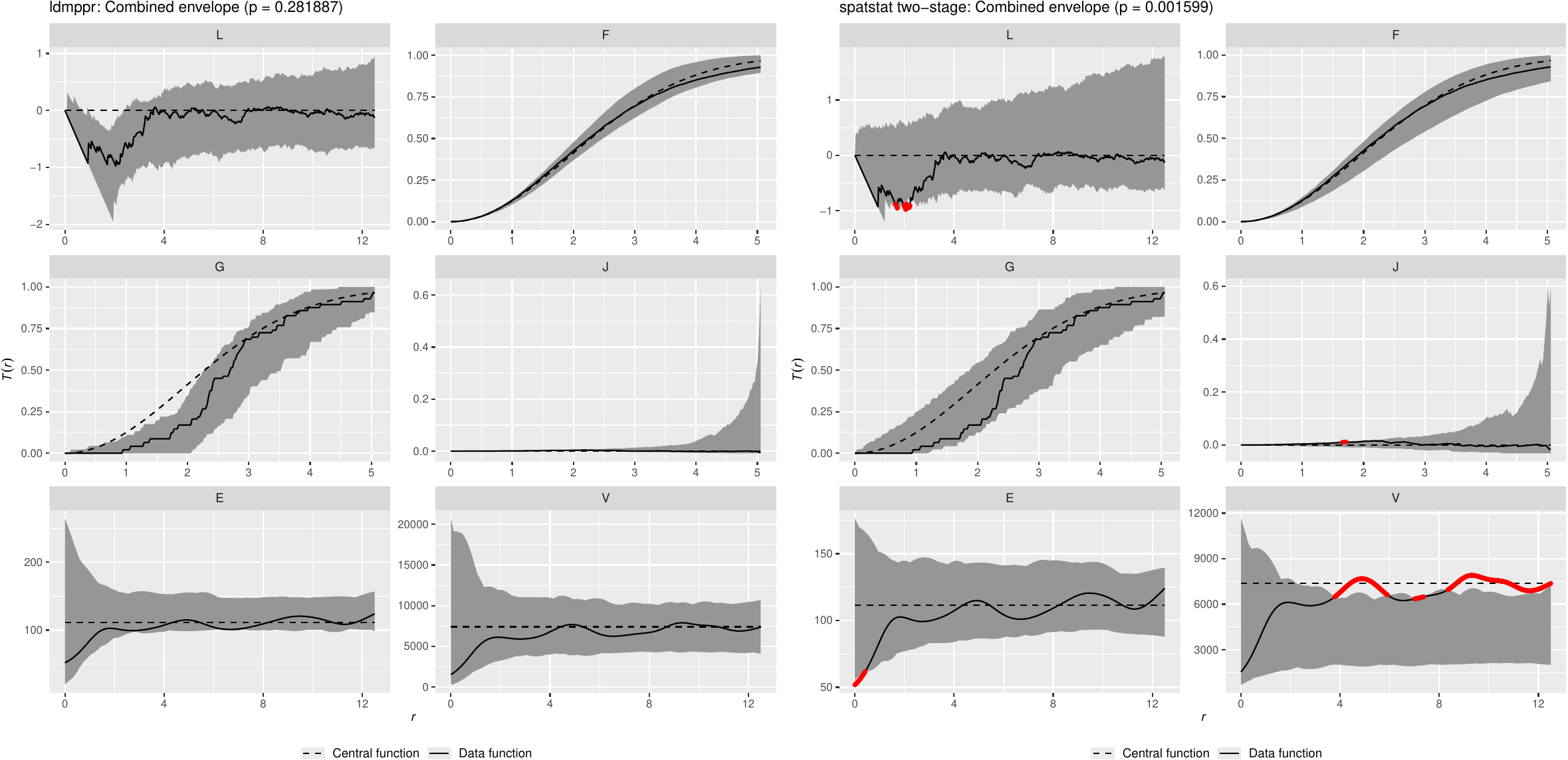} \end{center}

\end{landscape}

\begin{center}\includegraphics[width=1\linewidth]{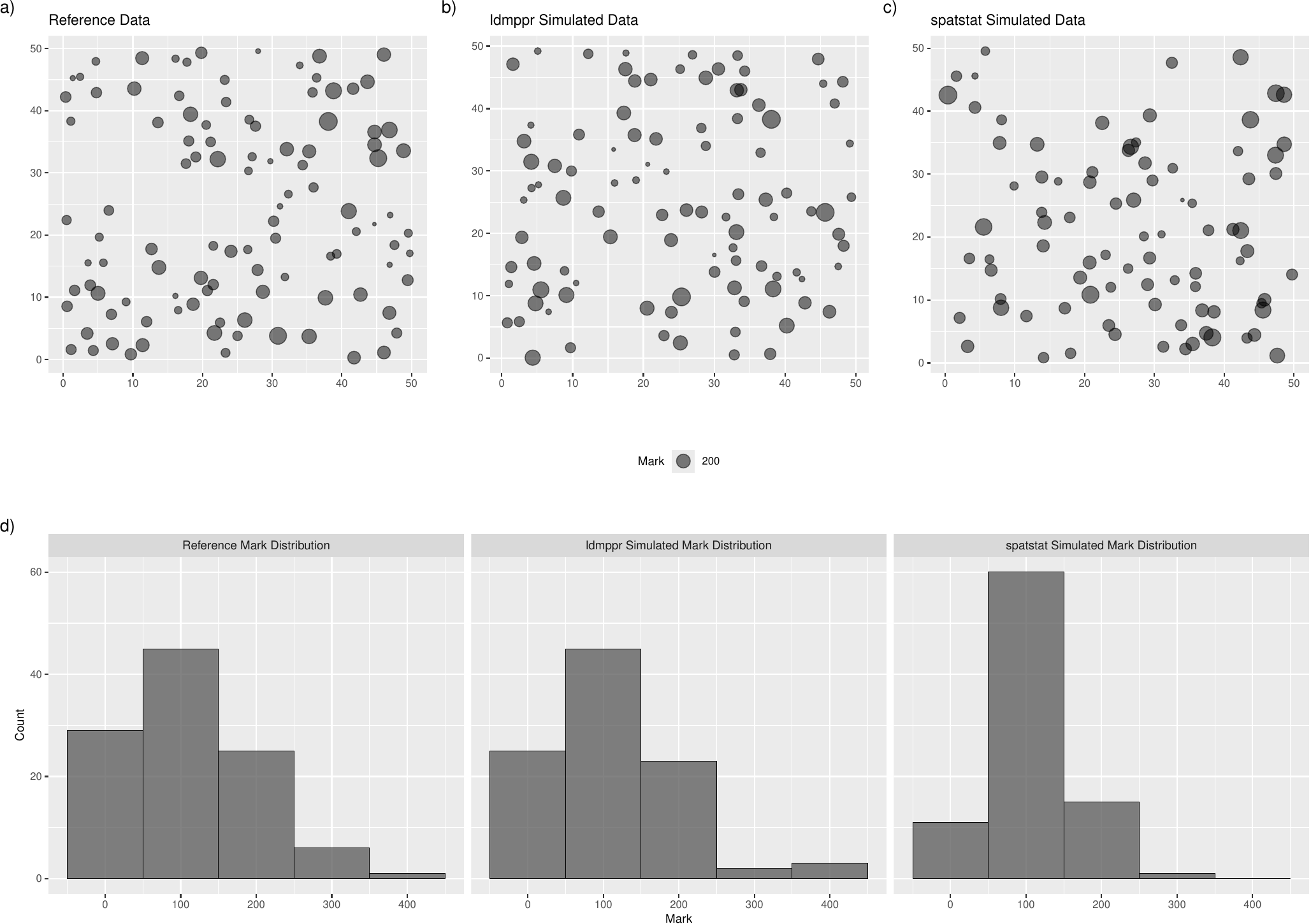} \end{center}

This baseline provides useful context but is not a fully comparable
benchmark for the mechanistic \texttt{ldmppr} workflow. In particular,
temporal/self-correcting dynamics are not estimated in a single unified
model, and mark generation is modularized as a second stage. We
therefore interpret these results as contextual sensitivity evidence
rather than a definitive head-to-head comparison.

\address{%
Lane Drew\\
Colorado State University\\
Department of Statistics, Colorado State University\\ 851 Oval Dr, Fort Collins, CO 80524\\
\url{https://lanedrew.com/}\\
\textit{ORCiD: \href{https://orcid.org/0009-0006-5427-4092}{0009-0006-5427-4092}}\\%
\href{mailto:lane.drew@colostate.edu}{\nolinkurl{lane.drew@colostate.edu}}%
}

\address{%
Andee Kaplan\\
Colorado State University\\
Department of Statistics, Colorado State University\\ 851 Oval Dr, Fort Collins, CO 80524\\
\url{https://andeekaplan.com/}\\
\textit{ORCiD: \href{https://orcid.org/0000-0002-0912-0225}{0000-0002-0912-0225}}\\%
\href{mailto:andee.kaplan@colostate.edu}{\nolinkurl{andee.kaplan@colostate.edu}}%
}

\end{article}

\end{document}